\pdfoutput=1
\documentclass[a4paper,fleqn,usenatbib]{mnras}

\usepackage{mathptmx}
\usepackage[T1]{fontenc}
\usepackage{ae,aecompl}
\usepackage[usenames,dvipsnames]{xcolor}

\usepackage{graphicx}	% Including figure files
\usepackage{amsmath}	% Advanced maths commands
\usepackage{amssymb}	% Extra maths symbols

\title[BAO from voids]{Measuring Baryon Acoustic Oscillations from the clustering of voids }

\author[Yu Liang et al.]{
Yu Liang$^{1}$\thanks{E-mail: liangyu13@mails.tsinghua.edu.cn (YL)},
Cheng Zhao$^{1}$,
Chia-Hsun Chuang$^{2}$,
Francisco-Shu Kitaura$^{2}$,
\and Charling Tao$^{3,1}$
\\
$^{1}$Tsinghua Center for Astrophysics (THCA) \& Department of Physics, Tsinghua University, Beijing 100084, China\\
$^{2}$Leibniz-Institut f\"ur Astrophysik Potsdam (AIP), An der Sternwarte 16, D-14482 Potsdam, Germany\\
$^{3}$CPPM, Universit\'{e} Aix-Marseille, CNRS/IN2P3, Case 907, 13288 Marseille Cedex 9, France
}

\newcommand{\lsim}{\mbox{${\,\hbox{\hbox{$ < $}\kern -0.8em \lower 1.0ex\hbox{$\sim$}}\,}$}}
\newcommand{\gsim}{\mbox{${\,\hbox{\hbox{$ > $}\kern -0.8em \lower 1.0ex\hbox{$\sim$}}\,}$}}

\date{Accepted XXX. Received YYY; in original form ZZZ}

\pubyear{2015}

\begin{document}
\label{firstpage}
\pagerange{\pageref{firstpage}--\pageref{lastpage}}
\maketitle

\begin{abstract}
We investigate the necessary methodology to optimally measure the baryon acoustic oscillation (BAO) signal, from voids based on galaxy redshift catalogues. 
To this end, we study the dependency of the BAO signal on the population of voids classified by their sizes. 
We find for the first time the characteristic features of the correlation function of voids including the first robust detection of BAOs in mock galaxy catalogues. These show an anti-correlation around the scale corresponding to the smallest size of voids in the sample (the void exclusion effect), and  dips at both sides of the BAO peak, which can be used to determine the significance of the BAO signal without any priori model.
Furthermore, our analysis demonstrates that there is a scale dependent bias for different populations of voids depending on the radius, with the peculiar property that the void population with the largest BAO significance corresponds to tracers with approximately zero bias on the largest scales.
We further investigate the methodology on an additional set of 1,000 realistic mock galaxy catalogues reproducing the SDSS-III/BOSS CMASS DR11 data, to control the impact of sky mask and radial selection function. Our solution is based on generating voids from randoms including the same survey geometry and completeness, and a post-processing cleaning procedure in the holes and at the boundaries of the survey.
The methodology and optimal selection of void populations validated in this work have been used to perform the first BAO detection from voids in observations, presented in a companion paper. 
\end{abstract}


%%%%%%%%%%%%%%%%%%%%%%%%%%%%%%%%%%%%%%%%%%%%%%%%%%

%%%%%%%%%%%%%%%%% BODY OF PAPER %%%%%%%%%%%%%%%%%%
\section{Introduction} 
%\label{sec:intro}

Baryonic Acoustic Oscillations (BAO) have been evidenced about ten years ago in the SDSS DR3
Luminous Red Galaxy sample \citep[][]{2005ApJ...633..560E} and in the 2dF survey \citep[][]{CPP05}. This was a smoking gun confirmation of the $\Lambda$CDM cosmology, after the Cosmic Microwave Background (CMB) radiation discoveries \citep[][]{WMAP103,WMAP711,WMAP913,PLANCKBAO14}, and the evidence for  an accelerated expansion of the Universe with Supernovae Type Ia (SNIa) \citep[][]{1998ApJ...507...46S,1998AJ....116.1009R,1998Natur.391...51P}.
The BAO signal has been improved by further SDSS measurements and other galaxy redshift surveys  \citep[][]{EZH05,PRE10,BKB11,BBC11,wigglez2010,boss2011,AAB14}. Recently BAOs were also detected in the Lyman alpha forest  \citep[][]{BDR13,SIK13,DBB15}. BAOs have become an incomparable tool for measuring angular diameter distances in the Universe, associated with the CMB signal from the time of recombination. BAO combined with CMB is more powerful today than the combination of SNIa and CMB to constrain cosmological parameters \citep[][]{ABB14} and there will be many surveys including BAO measurements as an integral part of their science goal, such as the the DESI/BigBOSS \citep[][]{bigboss2011}, the DES \citep[][]{des2013}, the LSST  \citep[][]{lsst2012}, the J-PAS \citep[][]{jpas2014}, the 4MOST \citep[][]{4most}, or the EUCLID survey \citep[][]{euclid2009}.

The first cosmic void, the so-called giant Bo{\"o}tes void, was detected by \citet{KOS81}. Since then there are more evidence of the voids that exist as one of the cosmic web structures. Cosmic voids are classified based on galaxy distribution of the surveys  \citep[][]{LGH86, VGP94, EP97, MAE00, PB02, CCG04, HV04, PBP06, CCW05, NKH14, K09, JKL10, PWJ11, VBT12, PVH12, SLW12, NH14, SLW14, BKH15} or dark matter density filed of the numerical simualtions \citep[][]{2005MNRAS.360..216C, 2006MNRAS.367.1629S, 2007MNRAS.380..551P, 2007MNRAS.375..489H, 2008MNRAS.386.2101N, 2009MNRAS.396.1815F, 2012MNRAS.425.2049H, 2013MNRAS.429.1286C}.

Many efforts are made to use cosmic voids as a cosmological probe to study the physics of the universe. 
Void probability distribution function and their cumulative void number density can be used to constrain $\sigma_8$ and $\Omega_m h$ \citep[][]{BPP09}. The void statistics has been studied by many work \citep[e.g.][]{W79,PP86,B90,EEG91,BL02}
Recently the shape of voids is used to constrain dark energy  \citep[][]{PL07,LW10,2015arXiv150307690P}. In particular, they can be used to test dynamical dark energy \citep{B12}, coupled dark energy \citep{L11},  and modified gravity \citep{MS09,LZK12,CC13,LCC15}.
They can also be used to measure the Sachs Wolfe effect  \citep{GNS08,ILD13,CLC14,HNG15,PLANCKISW14}.

Cosmic voids can potentially better trace the conditions of the primordial Universe than galaxies,
since their centres are less affected by nonlinear gravitational pull, as they expand in a nearly isotropic way \citep[see][and references therein]{SW04}. 
However, the number of voids, as is usually defined, is too small to allow a detection of the BAO peak (e.g., \citealt{CJS15}).
%\citep[][]{2015arXiv150307690P}.

We study the clustering of the troughs of the density field as proposed for the first time in Kitaura et al (companion paper). Taking into consideration the  subvoids in addition to the disjoint (parent) voids overcomes the issue of low void statistics, and allows statistically significant studies of clustering at the BAO scales, which was not possible previously \citep[see e.g.][]{PPH06,VBT12,CC13}.  In fact, the number of tracers can increase by about two orders of magnitude from considering only disjoint voids to including overlapping sub-voids (see Zhao et al.; companion paper). Whether this definition permits us to study BAOs from expanding empty regions in the Universe needs to be verified, which is the aim of this work.
We have provided the algorithm to obtain estimates of the troughs of the density field \citep[\textit{voids-in-voids}, according to the terminology used in ][]{SW04}, based on the empty circumspheres constrained by tetrahedra of galaxies through Delaunay triangulation (for details see Zhao et al.; companion paper).
Nevertheless, this approach can be contaminated by groups of galaxies (\textit{voids-in-clouds}), which are anti-correlated to the troughs.
A thorough analysis of the BAO signal from such troughs of the density fields robustly determining the signal-to-noise ratio for different populations of voids needs to be done. 

In this paper, we study in particular, the dependencies of the BAO signal-to-noise and clustering bias to the void radius. We compute autocorrelation functions and cross-correlation functions of the voids with different radius bins or cuts, and seek the  population of voids  with the highest BAO signal detection. 
We consider a large set of complete halo catalogues constructed with the \textsc{patchy} code \citep[][]{2014MNRAS.439L..21K,2015MNRAS.450.1836K}.
In addition, and to enable a robust measurement of BAOs from voids in observational data, we investigate the methodology used in the galaxy clustering analysis generalised to voids clustering. To this end, we consider a large set of accurate lightcone  mock galaxy catalogues resembling the BOSS CMASS DR11 clustering and survey geometry \citep[][]{2015arXiv150906400K}, which have been calibrated based on a reference catalogue applying halo abundance matching to the BigMultiDark N-body simulation \citep[][]{2015arXiv150906404R}.

This paper is organized as follows. 
We first describe in \S \ref{sec:dive} the void finding algorithm used in this study. Then we analyse the measurement of BAO from voids  based on complete halo catalogues in cubical volumes in \S \ref{sec:BAOcomp}.
 Subsequently, we show how to optimally obtain the BAO signal from voids based on lightcone catalogues in \S \ref{sec:lightcone}. 
Finally we summarize and conclude in \S \ref{sec:summary}.

\section{DIVE: Delaunay trIangulation Void findEr}
\label{sec:dive}

The \textsc{dive} algorithm  has been introduced in Zhao et al. (companion paper).
We define voids as the empty circumspheres constrained by tetrahedra of galaxies. These are obtained by Delaunay Triangulation (DT) applied to 3-D spatial distribution of objects.
In particular, \textsc{dive} relies on the publicly available {Computational Geometry Algorithms Library}\footnote{\url{http://www.cgal.org}} \citep*[\textsc{cgal},][]{cgal:eb-15b}.
The centres of the spheres define the void position.
Studies of the DT void properties used \textsc{patchy} mocks \citet{2015arXiv150906400K} are presented in Zhao et al. 2015. According to this study, there are  two different classes of DT voids, which correspond to groups (\textit{voids-in-clouds}) and troughs (\textit{voids-in-voids}).  Small voids have a high probability of residing in dense regions. They mainly  trace the quartets of galaxies, corresponding to \textit{voids-in-clouds} type voids.
 Large voids are more likely to trace underdense expanding regions, and correspond to \textit{voids-in-voids} type voids. Voids, as usually defined in the literature \citep[see e.g.][]{PPH06,VBT12,CC13}, correspond to a subclass of our voids, which do not overlap with each other, and thus we dub them ``disjoint'' voids. In this study, we allow the voids to overlap as proposed by Kitaura et al.; companion paper, to maximize the information obtained from DT voids tracing the troughs of the density field, which permits us to extract useful measurements, such as the BAO from void clustering.

\section{BAO from voids in complete mock halo catalogues}
\label{sec:BAOcomp}

The first aim of our study is to estimate the significance of the BAO measurements from voids obtained from complete halo catalogues. As we need large sets of mock catalogues which cover huge cosmic volumes, we rely on accurate and efficient mock generation methods described in \S \ref{sec:galaxy_catalog}. We then construct the void catalogues \S \ref{sec:voidscc} and compute the two-point correlation functions \S \ref{sec:bao_box}. Then we  analyse the characteristic clustering of voids and introduce a model independent signal-to-noise estimator in \S \ref{sec:ston}. Based on that we estimate the optimal radius cut, which defines the population of voids with the highest BAO signal  \S \ref{sec:optr}. we analyse the properties of the sub-populations using a cross-correlation study \S \ref{sec:cross}.

\subsection{Input data: Halo catalogues from \textsc{patchy} simulations in cubical volumes}
\label{sec:galaxy_catalog}
%\subsection{\textsc{patchy} simulation boxes}

We start with halo mock catalogues resembling the clustering of BOSS Luminous Red Galaxies with number density around $3.5\times10^{-4}\,h^3\,\mathrm{Mpc}^{-3}$, at a mean redshift of  $z\simeq0.56$ in cubical volumes of 2.5\,$h^{-1}$Gpc side.  
These are constructed with the PerturbAtion Theory Catalogue generator of Halo and galaxY distributions \citep[\textsc{patchy},][]{2014MNRAS.439L..21K}, which includes an explicit Eulerian nonlinear and stochastic bias description.
The input parameters of the \textsc{patchy} mocks are calibrated \citep[][]{2015MNRAS.450.1836K} with the Bound Density Maximum (BDM, including sub-halos) halo catalogue of the BigMultiDark $N$-body simulations \citep[][]{Klypin2014} performed with the cosmological parameters $\Omega_{\rm M} = 0.307115, \Omega_b = 0.048206, \sigma_8 = 0.8288, n_s = 0.96$, which is the Planck $\Lambda$CDM cosmology, and the Hubble parameter $H_0 \equiv 100\,h\,\mathrm{km}\,\mathrm{s}^{-1}\mathrm{Mpc}^{-1}$ is given by $h = 0.6777$.
The accuracy of the \textsc{patchy} mocks in terms of two and three point statistics, both in configuration and in Fourier space, with and without redshift space distortions (RSDs), was demonstrated to be very high, compared to reference $N$-body simulations  \citep[see][]{2015MNRAS.446.2621C}.
In particular, we generate 100 catalogues with varying seed initial conditions, in real and redshift space,  with the distant observer approximation, and within cubical volumes of 2.5\,$h^{-1}$Gpc side. This permits us to obtain robust error estimates.

\begin{figure}
\begin{center}
\includegraphics[width=.47\textwidth]{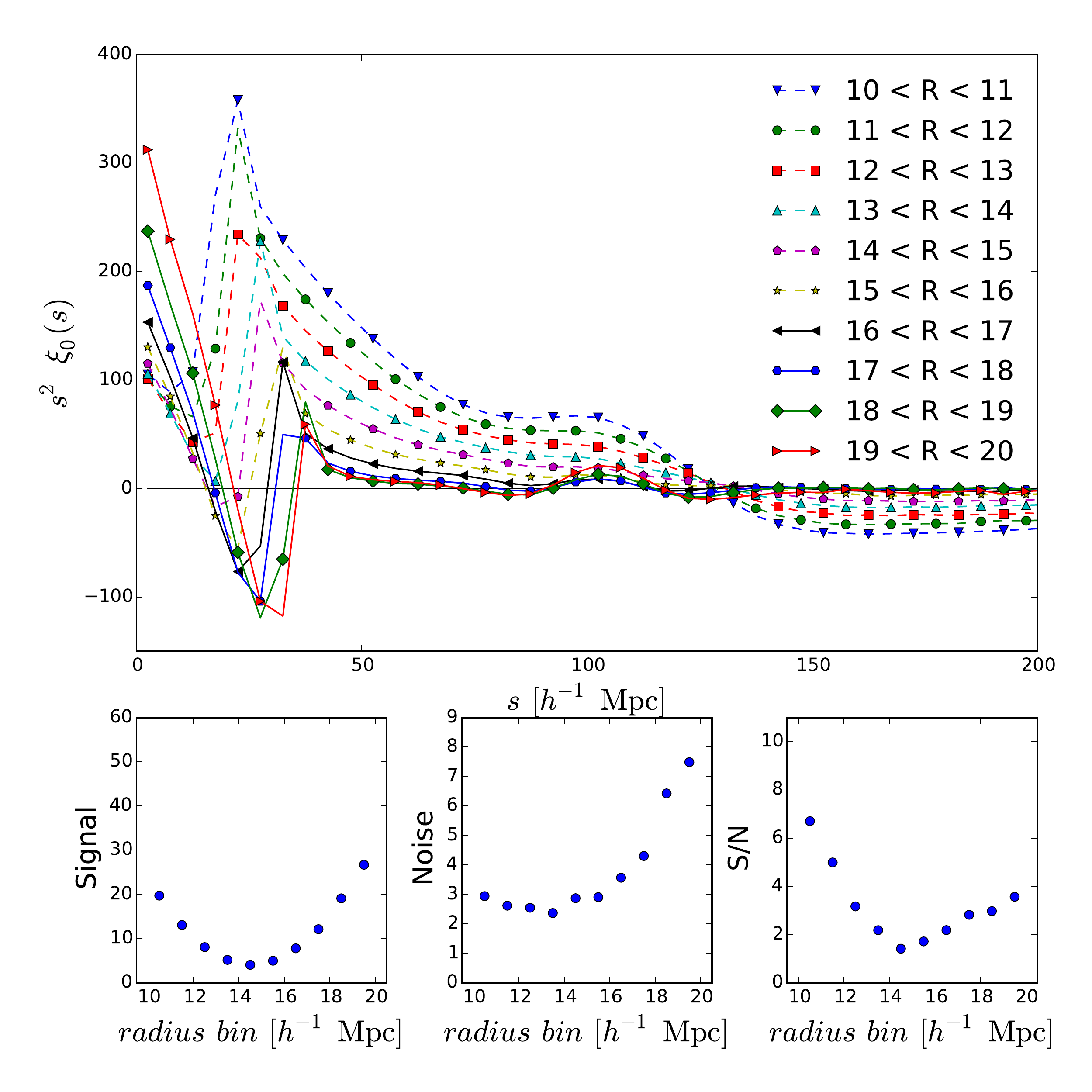}
\includegraphics[width=.47\textwidth]{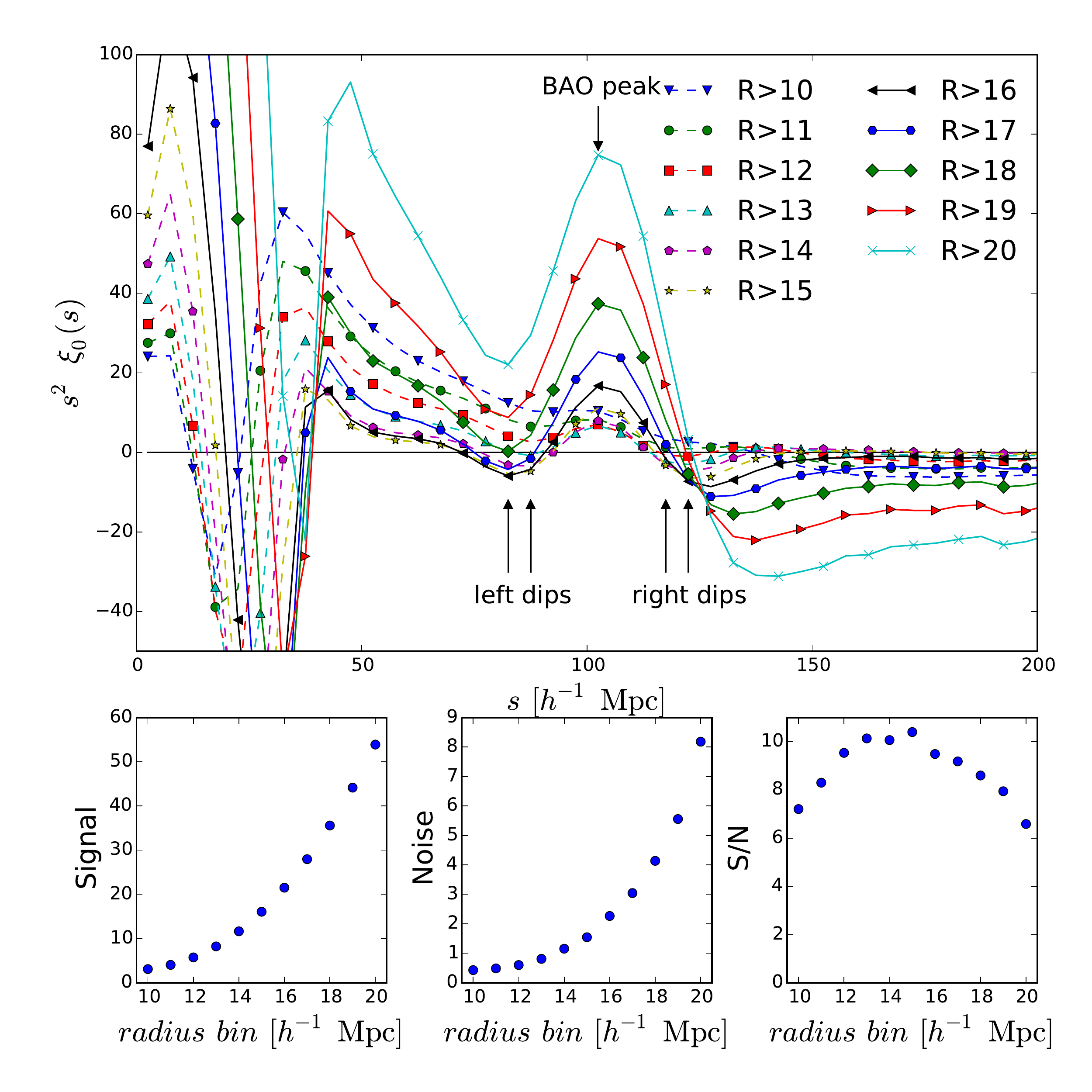}
\caption{Real-space two-point correlation function from voids based on halo catalogues in cubical volumes. The upper panel shows the results of voids with radius bins $R_1<R<R_2$; the lower panel shows the results from voids with radius cuts, $R>R_{\rm cut}$. The radius is in units of $h^{-1}$Mpc. Each panel includes sub-panels showing the signal $S$, noise $N$, and signal-to-noise ratio $S/N$.}
\label{fig:CFrealspace}
\end{center}
\end{figure}

\begin{figure}
\begin{center}
\includegraphics[width=.47\textwidth]{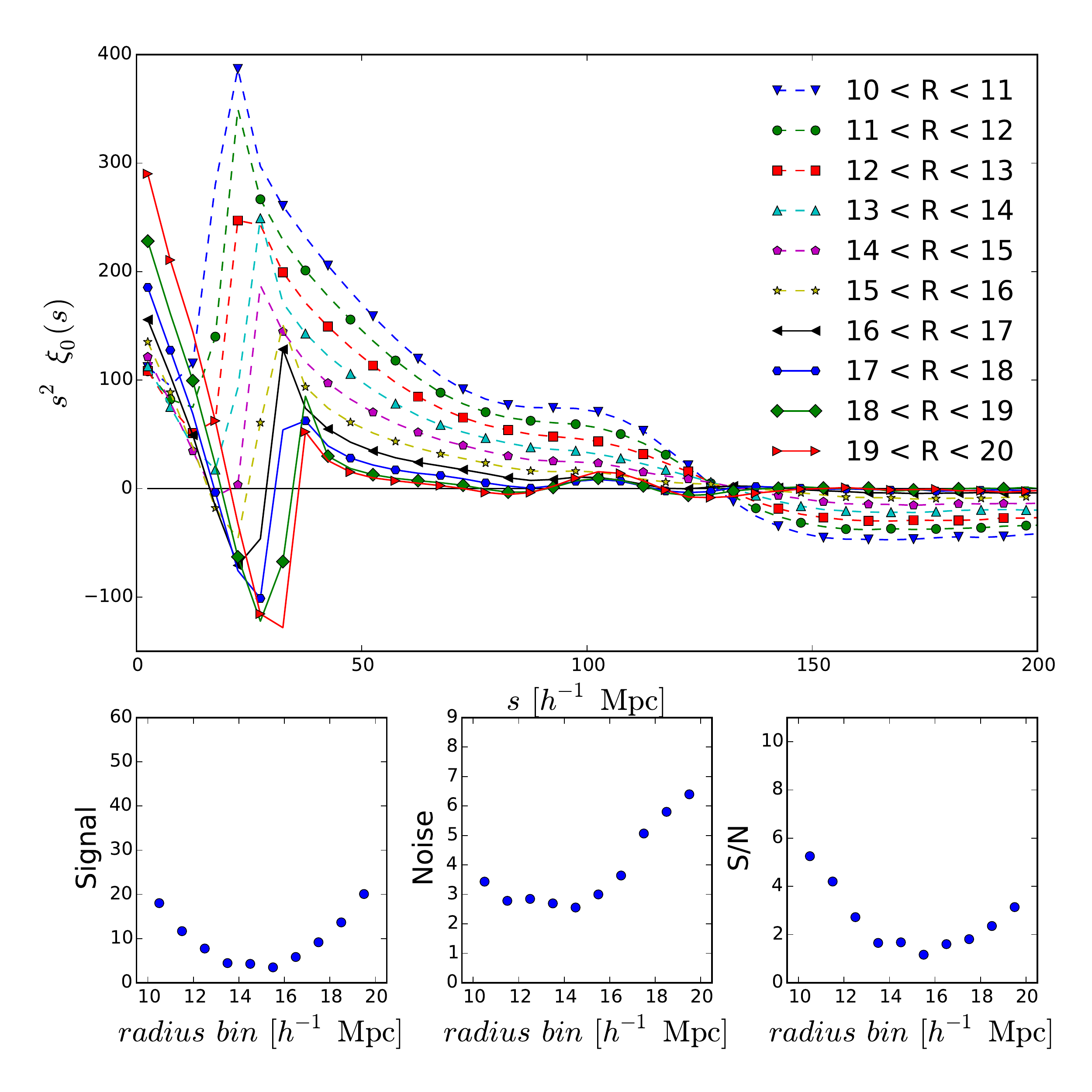}
\includegraphics[width=.47\textwidth]{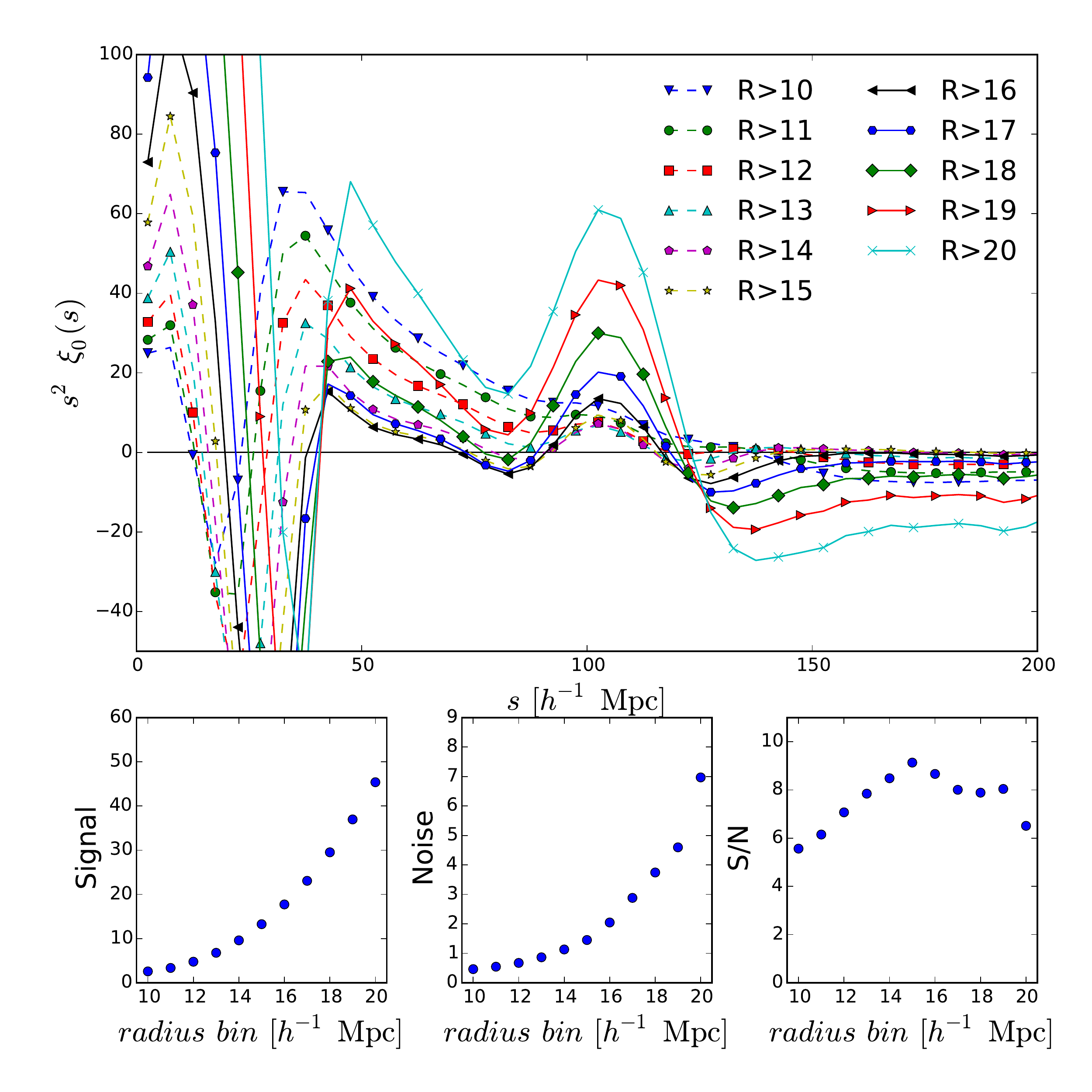}
\caption{Same as Fig.~\ref{fig:CFrealspace} but including redshift space distortions.}
\label{fig:CFzspace}
\end{center}
\end{figure}

\subsection{Construction of void catalogs for halo catalogues in cubical volumes}
\label{sec:voidscc}

We apply the \textsc{dive} algorithm to both sets of halo catalogues in real and in redshift space to obtain the corresponding void catalogues. In this work, we consider DT voids with the radius down to 10 $h^{-1}$ Mpc, As demonstrated in Zhao et al. (companion paper) smaller void sizes are dominated by \textit{voids-in-clouds}, which are not the troughs of the density field. We note that the redshift space distortion effect for DT voids is different from that of galaxies. 
{ Voids cannot be treated as point-like objects within large-scale structure analysis, in contrast to galaxies or haloes. 
In our work we are not moving voids from real to redshift space, as we do with haloes, but compute the distribution of voids in each space according to the corresponding distribution of haloes. This implies a crucial difference between haloes (or galaxies) and voids. 
While RSDs cause virtual displacements to haloes (or galaxies) along the line-of-sight,  their impact on voids does not only displace them,  but could change their size, or even make them  disappear/appear when a galaxy moves in/out of the circumsphere constrained by the tetrahedron of galaxies.   We will study the effects of RSD on DT voids in detail in future work, and restrict the discussion  in this paper  to the impact on the measurement of the BAO signal. }

\subsection{Two-point correlation function estimator for complete catalogues in cubical volumes}
\label{sec:bao_box}

To measure the BAO from voids, we compute the two point correlation function $\xi(s)$ from the void catalogues, where $s$ is the separation between a pair of void centres. Throughout this paper, our correlation function plots are modulated by the squared distance, $s^2 \,\xi(s)$, to visually enhance the BAO signal.

For the simulated boxes with periodic boundary condition, the correlation function is computed following the  \citet{1974ApJS...28...19P} estimator,
\begin{equation}
\xi (s)=\frac {DD(s)} {RR(s)} -1\,,
\end{equation}
where the $DD$ term is the pair count within a given bin of separation from s$_{\rm min}$ to s$_{\rm max}$ normalised by the total number of pairs. The value of the $RR$ term can be analytically computed through the following expression
\begin{equation}
RR(s) = \frac{4\pi}{3}  \frac {s_{\rm max}^3 - s_{\rm min}^3}{2V}\,,
\end{equation}
where $V$ is the volume of the box.

\subsection{Characterisation of the correlation function from voids: model independent BAO signal-to-noise estimator}

\label{sec:ston}

%The correlation function from voids has very characteristic features.
All void correlation functions show an anti-correlation spike with its minimum around the scale corresponding to the size of the void (twice the radius bin: see upper panel in Fig.~\ref{fig:CFrealspace}, or twice the radius cut: see lower panel in Fig.~\ref{fig:CFrealspace}). The position of the dip is about twice  the minimum void radius.  This is due to the void exclusion effect \citep{Hamaus:2013qja}. 

A peak at the BAO scale of $\sim$102.5  $h^{-1}$ Mpc can be identified for radii above 16 $h^{-1}$ Mpc. The most prominent additional feature we find (for radius bins $17<R<18$ till $19<R<20$ $h^{-1}$ Mpc, and radius cuts  $R>13$  $h^{-1}$ Mpc) are two dips, one at scales smaller  (left of BAO peak: $\sim$85  $h^{-1}$ Mpc) and one at scales larger (right of BAO peak: $\sim$120  $h^{-1}$ Mpc)  than the BAO peak ($\sim$102.5  $h^{-1}$ Mpc). This characteristic pattern is more pronounced than for the clustering of haloes, and permits us to define in a model independent way an efficient estimator of signal-to-noise ratio ($S/N$). 

Encouraged by the characteristic signal in the correlation function of voids, we define the signal
$S$ as
\begin{equation}
S\equiv \xi(r^{\rm BAO}) - \frac{\xi(r_{\rm 1}^{\rm dl})+\xi(r_{\rm 2}^{\rm dl})+\xi(r_{\rm 1}^{\rm dr})+\xi(r_{\rm 2}^{\rm dr})}{4}\,.
\end{equation}
For the cosmological parameters we are considering we find from our calculations, that it is appropriate to define: $r^{\rm BAO}=102.5$, $r_{\rm 1}^{\rm dl}=82.5$, $r_{\rm 2}^{\rm dl}=87.5$, $r_{\rm 1}^{\rm dr}=117.5$, and $r_{\rm 2}^{\rm dr}=122.5$  { { $h^{-1}$ Mpc}. 
The noise $N$ is then defined as the standard deviation of the signals measured from 100 void \textsc{patchy} mocks.
We show a model dependent signal-to-noise estimator extracted from mock catalogues in Kitaura et al. (companion paper). We leave an analytical modelling of the clustering of voids for future work.
The bottom sub-panels  in Figs.~\ref{fig:CFrealspace}, \ref{fig:CFzspace} and \ref{fig:CFlightcone} show the signal, noise, and signal-to-noise ratios, respectively,  for different radius bins or cuts.
BAO shifts with respect to the one from the underlying dark matter field are expected to be at the level of 0.3\% for different halo/galaxy types \citep[][]{2014MNRAS.442.2131A,2014arXiv1410.4684P}.  Since in this study we are not aiming at estimating the exact BAO peak position, but only its detectability, we postpone such a precision study for future work.

\subsection{Optimal void radius cut for BAO from complete catalogues}

\label{sec:optr}

From the previous calculations we find that the optimal void radius cut yielding the highest signal-to-noise ratio is around 15 $h^{-1}$ Mpc (see Fig.~\ref{fig:CFrealspace}). 
Interestingly the results from redshift space yield a similar radius cut (see Fig.~\ref{fig:CFzspace}).
However, the significance is reduced from about 10.4  to 9.3 $\sigma$ including RSDs.
The ``V'' shape in the signal-to-noise ratios from radius bins, or the equivalent inverted ``V'' shape from radius cuts, around the optimal values, indicate that there are two anti-correlated population of objects both with a BAO signal. 
%We also compute the 2PCF of voids from the boxes in redshift space as shown in Fig.~\ref{fig:CFzspace}. We observe that the optimal void radius cut increases to 15.5 Mpc/h, since the void radius statistically grow from real to redshift space (see Zhao et al. 2015).  
 
\begin{figure}
\begin{center}
\includegraphics[width=.47\textwidth]{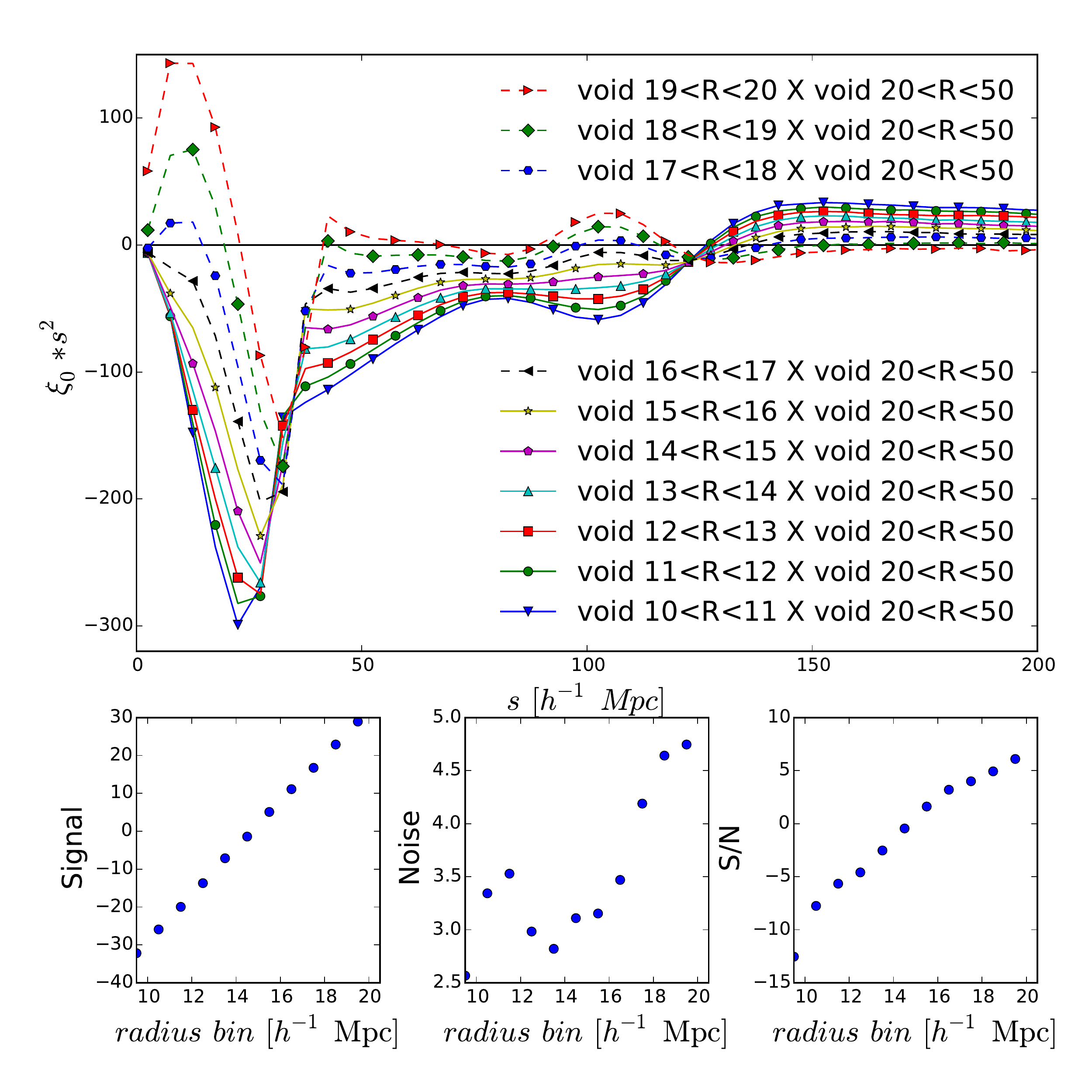}
\caption{Cross correlation functions between different radius bins and the reference void catalogue with $20<R<50$ in $h^{-1}$ Mpc. 
}
\label{fig:xcf}
\end{center}
\end{figure}

\subsection{Cross-correlation function analysis}
\label{sec:cross}

To verify the existence of two anti-correlated populations of voids, we compute the cross-correlation between different populations of voids. In particular, we choose as a reference the voids with radii larger than 20 $h^{-1}$ Mpc ($R_{\rm ref}>20$ $h^{-1}$ Mpc). This radius cut is large enough to ensure that the population of voids is completely dominated by \textit{voids-in-voids}, i.e. by {\it true} voids residing in expanding regions (see Zhao et al.; companion paper).
Thus, any positive cross-correlation between another population (with a radius cut $R<R_{\rm ref}$) and the reference population indicates that they are also contributing to enhance the BAO signal from \textit{voids-in-voids}. Nevertheless, our results show a complex scale dependent bias (Fig.~\ref{fig:xcf},). 
The BAO signal change the orientation at about 15 $h^{-1}$ Mpc because of the transition of the dominant void population from \textit{voids-in-voids} to \textit{voids-in-clouds} for smaller radii. This explains the dependence of the BAO $S/N$ on different radius cuts: the BAO signal is canceled out while including smaller voids. The amplitudes of the cross-correlation functions at large scales (i.e., $s$ > 150 $h^{-1}$  Mpc) present the relative linear biases with respect to the reference void sample (i.e., $R_{\rm ref} > h^{-1}$  20 Mpc). Note that the linear auto-correlation function is negative at scales larger than $s \sim$ 150  $h^{-1}$ Mpc, so that the relative linear bias is negative if the cross-correlation function is positive. We can see that only the radius bin $19 < R < 20$  $h^{-1}$ Mpc has positive relative bias with respect to our reference voids sample. It means that the void linear bias changes sign with radius $\sim$ 19  $h^{-1}$ Mpc, which is consistent with \citealt{Hamaus:2013qja}, even though the void definitions are not the same. Also note that the linear bias and BAO signal vanish with different void radius due to the non-linear bias.
Nevertheless, the bias of voids cannot be well modelled with a linear bias, and including population of voids which have apparently an opposite sign in the bias (from large scales) share however the same BAO peak orientation, and thus contribute to enhance the void BAO signal.

In Fig.~\ref{fig:xcf}, all the cross-correlation functions intersect at one point around 130 $h^{-1}$ Mpc which is related to the size of the particle horizon at matter-radiation equality, estimated for the present Planck values \citep[][]{1994ApJ...428..399K,2011arXiv1111.2889P}. 
%Standard CDM cosmology fixes the size to be inversely proportional to the matter density $(h^2\,\Omega_{\rm M})^{-1}$, as discussed, e.g., by Peacock (1999). The linear correlation function crosses zero at this scale. Sylos Labini et al. (2009) have discussed to measure it for large scale galaxy surveys since the scale, $r_c$, is not sensitive to galaxy bias. 
\citet[][]{2009A&A...505..981S} have discussed to measure the scale, $r_c$, where the galaxy correlation function turns from positive to negative.
But, it is difficult to measure this scale due to the systematics error and statistics uncertainty from observations. For voids clustering, the impact of observational systematics works differently so that one might observe $r_c$ in the void correlation function but not in the galaxy correlation function. However, we find that the uncertainty of the position of $r_c$ is large (see Fig.~\ref{fig:CFrealspace} and ~\ref{fig:CFzspace}) because of the non-linear bias previously discussed. Thus, it would be even more difficult to extract reliable cosmological information from the measurement of $r_c$ from the void correlation function.

\section{BAO from voids in  mock galaxy lightcone catalogues}
\label{sec:lightcone}

In this section we investigate the measurement of BAO from voids based on lightcone data that encodes redshift evolution, survey geometry, and selection effects. To this end, we rely on the accurate patchy mocks introduced in \S \ref{sec:mockslightcone}, and generalise the clustering analysis techniques from galaxies to voids \S \ref{sec:2ptlight}, additionally obtaining the optimal voids population radius cut for BAO analysis.

\subsection{Input data: Mock galaxy catalogues  from \textsc{MultiDark PATCHY} BOSS DR11 CMASS lightcones}
\label{sec:mockslightcone}

To validate our BAO measurement technique on observational data, we present the study on mock catalogues specifically generated for  BOSS CMASS DR11 galaxies.
The Sloan Digital Sky Survey (SDSS; \citealt{Fukugita:1996qt,Gunn:1998vh,York:2000gk,Smee:2012wd}) mapped over one quarter 
of the sky using the dedicated 2.5 m Sloan Telescope \citep{Gunn:2006tw}.
The Baryon Oscillation Sky Survey (BOSS, \citealt{Eisenstein:2011sa, Bolton:2012hz, Dawson:2012va}) is part of the SDSS-III survey. 
It collected the spectra and redshifts for 1.35 million galaxies, 230,000 quasars and 
100,000 ancillary targets. The final raw data, data release 12 (DR12; \citealt{Alam:2015mbd}), has been made publicly available\footnote{http://www.sdss3.org/}.
CMASS samples are selected with an approximately constant stellar mass threshold \citep{Eisenstein:2011sa}; 
The details of generating this sample are described in \cite{Reid:2015gra}. The mock catalogues reproducing the clustering of these objects are presented in \citep[][]{2015arXiv150906400K} , which are calibrated with the BigMultiDark N-body simulations using the Halo Abundance Matching (HAM) technique \citep[][]{2015arXiv150906404R}.
We restrict our study to mock galaxies resembling the SDSS-III BOSS DR11 CMASS catalogue in the redshift range $0.43<z<0.7$.

\subsection{Construction of void catalogs from incomplete lightcone catalogues}

\label{sec:void-lightcone}

To obtain the void catalogues from the input lightcone galaxy catalogues we present now a series of  steps which takes care of the survey geometry and selection function:
\begin{description}
%\item[1.]filter the galaxy mocks with the veto flag and fiber collision flag.
\item[1.] convert the angular coordinates (RA, DEC) and redshifts of the galaxies in a lightcone catalogue to cartesian coordinates $(x, y, z)$ in comoving distances.
\item[2.] run the \textsc{dive} void finder to construct DT voids catalogues
\item[3.] convert the centre of each void from $(x, y, z)$ to (RA, DEC, redshift).
\item[4.] remove the voids which have their centres outside the survey unmasked regions\footnote{To apply the geometry, we use the code at \url{https://github.com/mockFactory/make_survey}} \citep[][]{2014MNRAS.437.2594W}.
\end{description}

Although the input galaxy catalogues are matching the survey geometry, the void centres identified by \textsc{dive} can be outside the survey. Step 4 is necessary to filter out these centres. Nevertheless, we cannot avoid the impact of the boundary on the voids inside the survey region. 
Since a void is determined by tetrahedra of galaxies, it could disappear if one of the 4 galaxies is located outside the mask. A new void could also appear if the sphere of 4 galaxies inside the survey region exclusively contains some galaxies outside the survey region. In this case the Delaunay condition would be artificially accomplished.  This boundary effect will change the void number density close to boundary and will bias the clustering measurements. Later, we will explain how we model this boundary effect in the random void catalogue to suppress the impact on the clustering measurements. 

\begin{figure}
\begin{center}
\begin{tabular}{c}
\includegraphics[width=.47\textwidth]{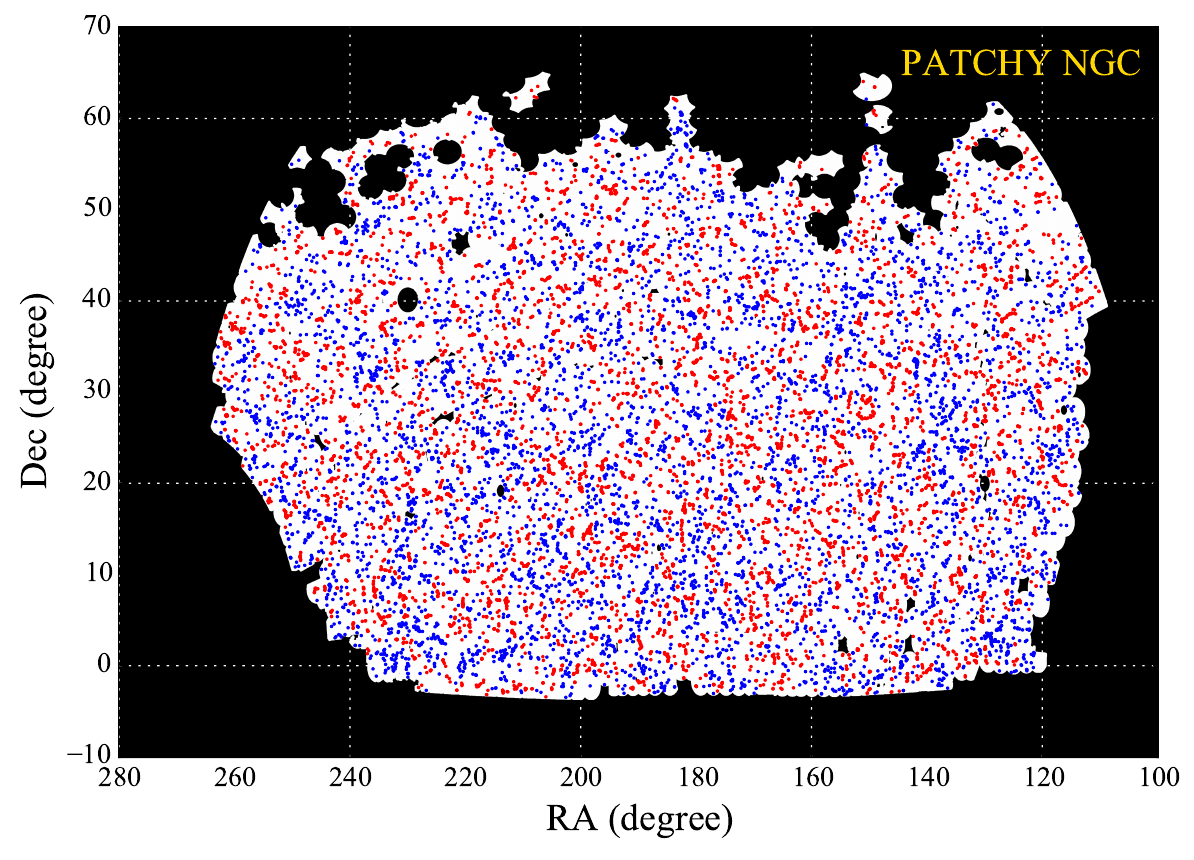}
%\put(-73,155.42){\footnotesize\color{Goldenrod} MD}
\end{tabular}
\caption{The angular distribution of \textsc{MultiDark PATCHY} BOSS NGC CMASS DR11 mock galaxies (blue) and void centres (red) in a redshift slide, 0.498 < z < 0.5. We show only the voids with radius larger than 16 $h^{-1}$ Mpc.}
\label{fig:cmass_sky_ngc}
\end{center}
\end{figure}

\subsection{Calculation of two-point correlation functions from incomplete lightcone catalogues}

\label{sec:2ptlight}

\subsubsection{Two-point correlation function estimator for lightcone catalogues}
We compute the two-point correlation functions from the 1,000 \textsc{patchy} lightcone mocks using the   \citet[][]{1993ApJ...412...64L} estimator:
\begin{equation}
\xi (s)=\frac {DD(s)-2DR(s)+RR(s)} {RR(s)} \,,
\end{equation}
where $DD$, $DR$, and $RR$ correspond to the normalized data-data, data-random, random-random pair counts. We use the void catalogue constructed  from 1000 DR11 CMASS mocks galaxy catalogues as described in \S \ref{sec:void-lightcone}. As opposed to the previous study on complete cubical volumes, one needs now to use random catalogues to estimate $DR$ and $RR$ terms. The random catalogues cannot be constructed from the random galaxy catalogues since the voids radius distribution would be very different. We develop a method to construct the random void catalogues, as described below. 

\subsubsection{Construction of random void catalogues}

Our goal here is to generate random void catalogues which share the same number density of voids in both angular and radial directions, as the observed one. To this end, we apply the ``shuffling'' method,  to reproduce the radial selection function. This method randomly assigns the redshift with radius taken from a given observed void catalogue  to a random angular point. The catalogues of the random angular points are generated based on the survey masks. The same method was used by the BOSS collaboration to construct the random catalogues for the galaxy sample. However, we cannot use the survey masks used by galaxy catalogues directly because of the ``boundary effect'' for voids, caused by the galaxies ``outside'' the survey area as described in Sec.~\ref{sec:void-lightcone}.

%When applying a survey mask to a full sky void catalogue, removing the galaxies outside the survey area, the galaxies inside the survey area are not affected. 

% move the following above
%However, since a void is determined by tetrahedra of galaxies, a void could disappear if one of the 4 galaxies is located outside the mask. A new void could also appear if the sphere of 4 galaxies inside the survey region exclusively contains some galaxies outside the survey region. In this case the Delaunay condition would be artificially accomplished.  This boundary effect will change the void number density close to boundary and will bias the clustering measurements. 
Here, we demonstrate this effect before describing the steps we use to construct the proper random catalogues for DR11 mock void catalogues.
%To demonstrate this effect, 
We construct two sets of test void catalogues, based on real space  \textsc{patchy} boxes described in \ref{sec:voidscc}. The test void catalogues have the same geometry: $10<{\rm RA}<80$, $10<{\rm DEC}<50$ and $0.4<z<0.7$. We do not apply the selection function, so that the random catalogue is homogeneous in our test survey geometry.  
\begin{itemize}
\item Set 1, the sample not having boundary effects is generated by applying the survey geometry on the void catalogues in the real space \textsc{patchy} boxes. 
\item Set 2, the sample including boundary effects is generated by applying the survey geometry on the galaxy catalogues in the same real space \textsc{patchy} volumes and then running \textsc{dive} to find the voids in the same way we construct the lightcone void catalogue.
\end{itemize}
In other words, set 1 and 2 identify voids and apply geometry in different order. The expected correlation function of set 1 is the same as the one of the void catalogues in boxes. We apply the geometry to set 1 so that it will share the same cosmic variance as set 2.
We use 20 catalogues for each set, for demonstration purposes.  Fig.~\ref{fig:boundary} shows that the boundary effect boosts the correlation function. We do not show error bars, as the relative effect is reliable, since the two sets of void catalogues are constructed from the same realisations and have the same geometry.

\begin{figure}
\begin{center}
\includegraphics[width=.47\textwidth]{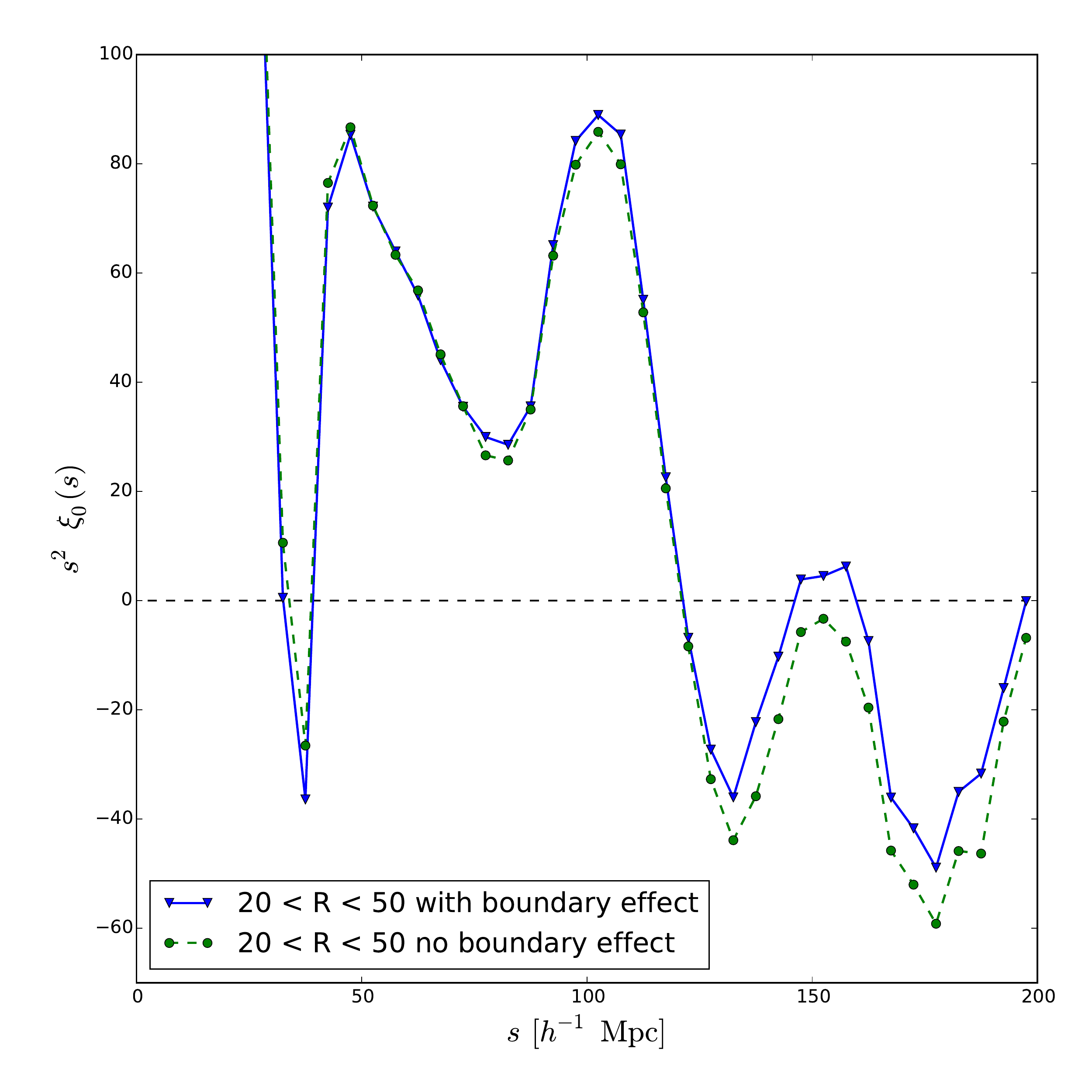}
\caption{Comparison of the void correlation function with (blue line) and without (green line) boundary effect.}
\label{fig:boundary}
\end{center}
\end{figure}

To reproduce the number density distribution including the boundary effect, we stack 100 mock void catalogues to construct the random angular point catalogue, given that the boundary effect is the same for the observed data and mocks. In addition, since the number densities are different at different redshifts, the boundary effect would also change. Therefore, we construct the random catalogues in narrower redshift bins, thereby minimizing the redshift dependency. 
We describe below the steps constructing the random catalogues to compute the correlation function  from  \textsc{patchy} lightcone mock void catalogues:  
\begin{description}
\item[1.] stack 100 void mock catalogues.
\item[2.] separate the stacked catalogue into two redshift bins, one from 0.45 to 0.55 and the other one from 0.55 to 0.65.
\item[3.] separate the two redshift bins in 5 radius bins ($16<R<17$, $17<R<18$, $18<R<19$, $19<R<20$, $20<R<50$ $h^{-1}$ Mpc), now the catalogue is divided in 10 subsamples.
\item[4.] split each subsample into two parts, RA, DEC on one side and $z$, R on the other. Shuffle the two parts separately and recombine them.
\end{description}

\begin{figure}
\begin{center}
\includegraphics[width=.47\textwidth]{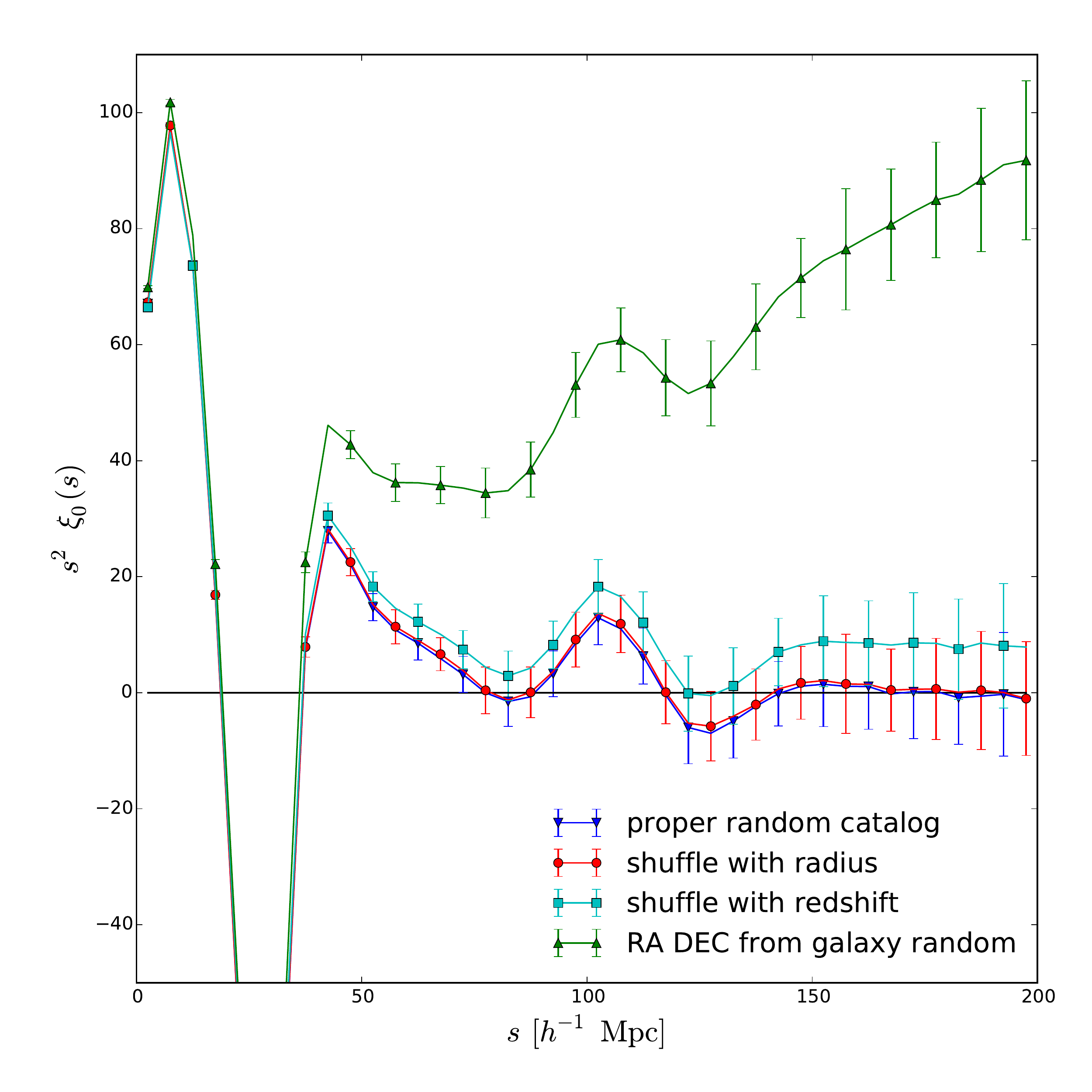}
\caption{ Comparison of the void correlation function with different random catalogs including 1. the proper random catalog (blue curve), 2. the random catalogue constructed without shuffling within radius bins (cyan curve), 3. the random catalog constructed without shuffling within redshift bins (red curve), and 4. random catalog constructed using mock galaxy randoms instead of stacking mocks as the angular random positions.}
\label{fig:shuffletest}
\end{center}
\end{figure}

Fig.~\ref{fig:shuffletest} shows the correlation functions using the proper random catalog constructed by the four steps described above, comparing with the ones constructed by skipping step 2 or step 3. In addition, we also take random angular positions from the galaxy random catalogues instead of stacking mocks and show the corresponding CF in Fig.~\ref{fig:shuffletest}. One can see that the CF is boosted if we ignore step 3 (shuffling within radius bins) or we use galaxy random catalogues. Our results are not sensitive to redshift dependency because we always keep the radius together with corresponding redshift as one part when constructed the random catalog. The boosting should due to the boundary effect described earlier (see Fig.\ref{fig:boundary}). The boosting is significant while using galaxy random catalogue because of the complex geometry of the real survey. In other words, it is critical to use the proper mock catalogues to build the random catalogue.

\subsection{Optimal void radius cut for BAO from BOSS CMASS DR11 mocks}

We compute the correlation functions from the first 100 \textsc{patchy} CMASS-NGC void catalogues and present them in Fig.~\ref{fig:CFlightcone} together with the BAO signal $S$, noise $N$, and signal-to-noise $S/N$ ratios.
The signal significance is lower, as expected, since the volume is roughly a factor 8 smaller  comparing to the cubical volumes studied in the previous section. 
The maximum $S/N$ is around radius cut of 16 $h^{-1}$ Mpc, but contrary to the case of complete real and redshift cubical volumes,   the $S/N$ does not show a decrease for larger radius cuts. The difference lies on the number density, which is now a function of redshift (i.e., radial selection function).
When the number density is lower, the radius of \textit{voids-in-clouds} would be larger. Therefore, when increasing the radius cut, we might throw away \textit{voids-in-clouds} in lower density redshift region and \textit{voids-in-voids} in higher density region. This could explain why the $S/N$ ratio is not sensitive to the cut for lightcone mocks. To further optimise the clustering measuerments, we should have a redshift dependent radius cut, which would depend on the number density at different redshifts. We leave such a study  for future work.
We find that the estimated $S/N$ for the \textsc{patchy} mocks is 2.35 for the BOSS CMASS-NGC with voids larger than 16 $h^{-1}$ Mpc.

\begin{figure}
\begin{center}
\includegraphics[width=.47\textwidth]{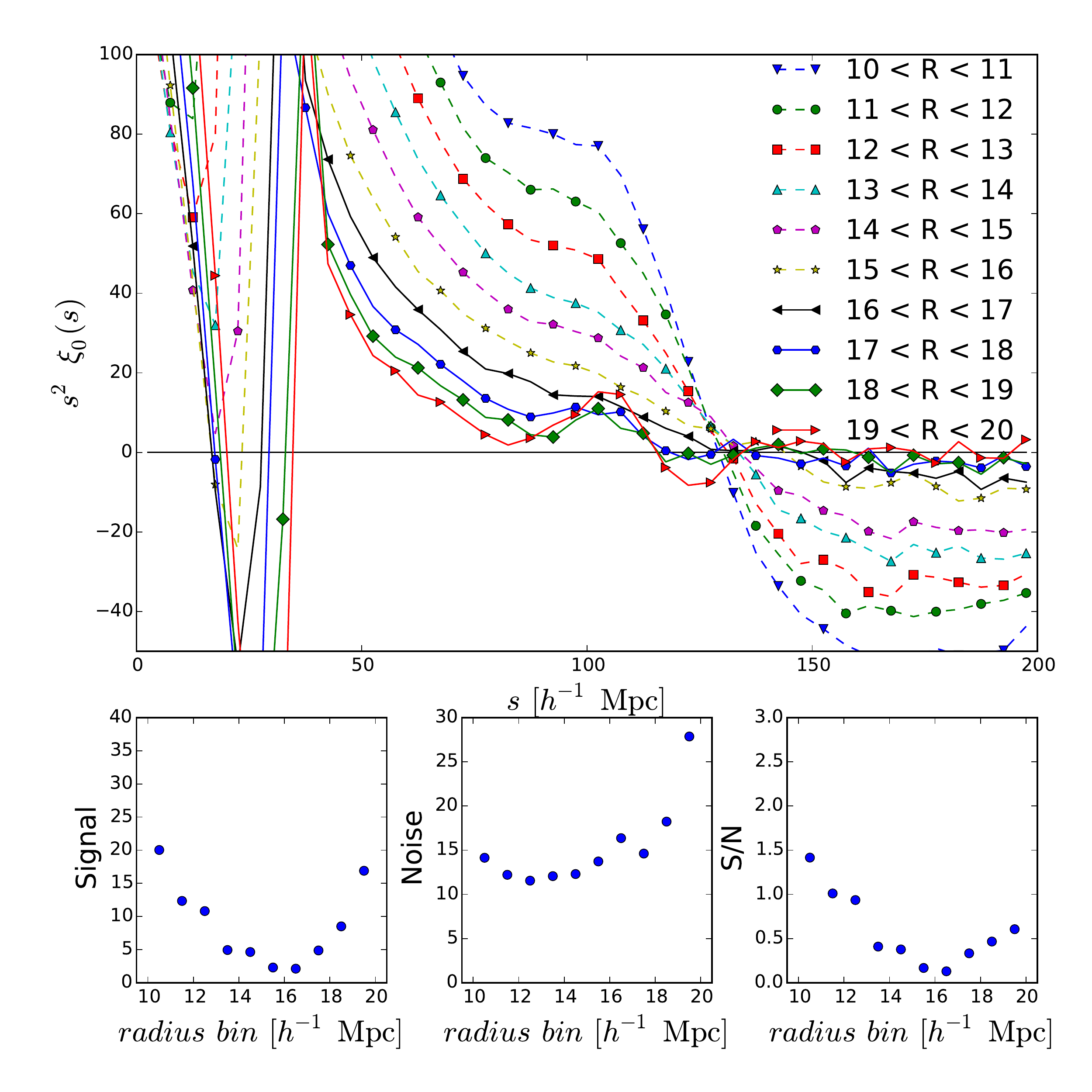}
\includegraphics[width=.47\textwidth]{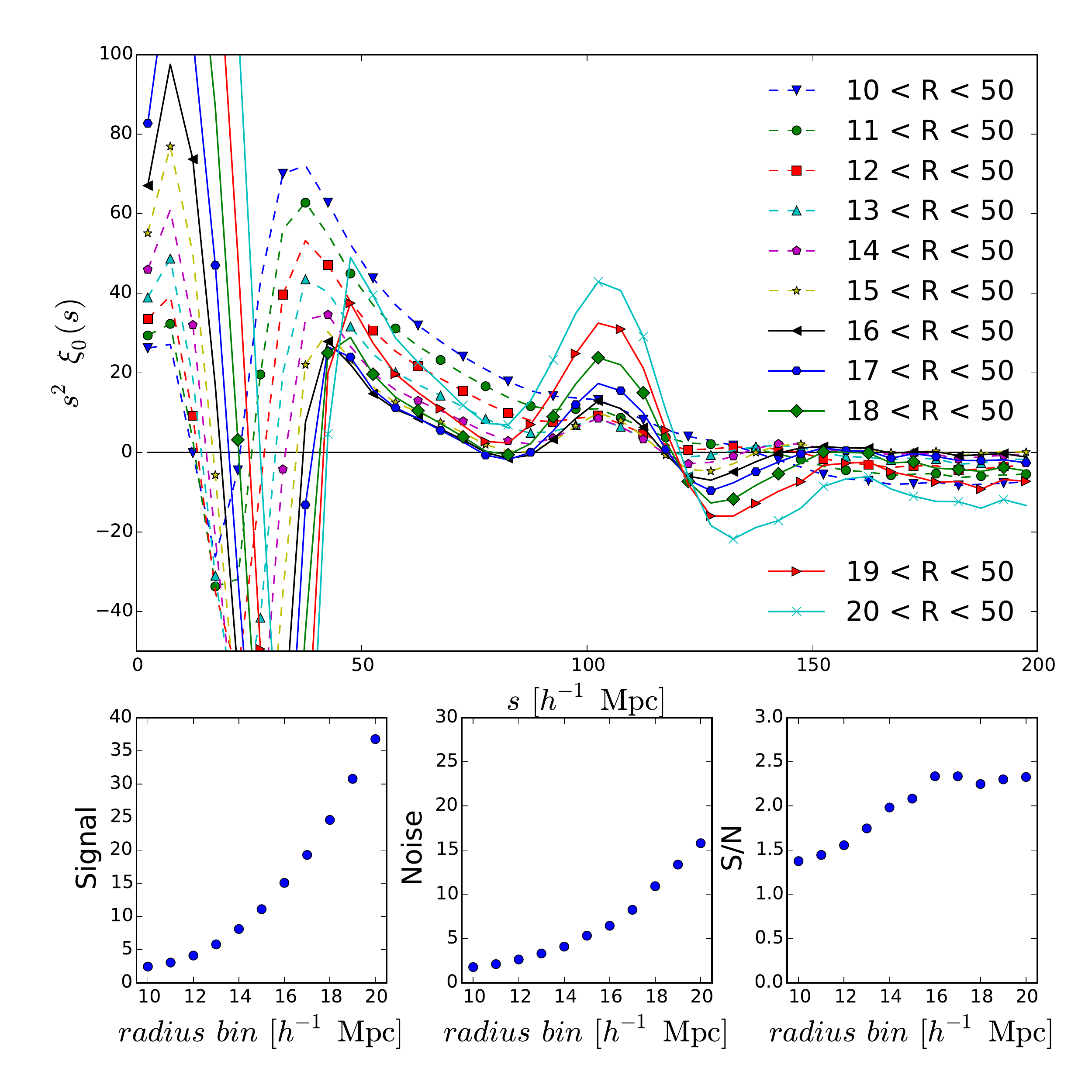}
\caption{Correlation functions measured from the first 100 \textsc{patchy} CMASS-NGC void catalogues with $R$ bins and $R>R_{\rm cut}$.}
\label{fig:CFlightcone}
\end{center}
\end{figure}

\section{Summary}

\label{sec:summary}

We have studied in this work the clustering of voids including sub-voids, and thus tracing the troughs of the density field. To this end, we have applied the novel \textsc{dive} void finding algorithm, which is particularly well suited for discrete distribution of objects (see Zhao et al; companion paper). This has permitted us to find for the first time the Baryon Acoustic Oscillations signal from voids in simulations. We have furthermore studied the signal-to-noise of BAO detections for different populations of voids, classified by their sizes. In addition, we have developed the necessary techniques to extract BAO signal from lightcone data including survey geometry and selection function effects.

Our study based on large sets of accurate mock catalogues demonstrates that the correlation function from voids, following our definition, has very characteristic features with dips around the BAO peak, which permits us to define a model independent signal-to-noise ratio.

Moreover, our results show that for BOSS CMASS Luminous Red Galaxy like objects at $z\simeq0.56$ with the number density of $3.5\times10^{-4}\,h^3\,\mathrm{Mpc}^{-3}$  the optimal void radius cut is 15$\pm1\,h^{-1}$ Mpc for both real and redshift space. We forecast signal-to-noise BAO detections of $>$9 $\sigma$ from complete volumes of $(2.5\,h^{-1}\,{\rm Gpc})^3$, which will be accesible with future surveys. 

Furthermore, our analysis demonstrates that there is a scale dependent bias for different populations of voids depending on the radius cut, with the peculiar property that the void population with the largest BAO significance corresponds to tracers with approximately zero bias on the largest scales.

The cosmological gain from using voids needs further investigation that will be presented in future work  (Chuang et al. in prep.).
Since voids are found based on the distribution of galaxies (or haloes), one may conclude that no additional information is present in the two-point correlation function of voids. However, voids are constructed upon tetrahedra of galaxies including information on higher order statistics. Presumably, the information from higher order statistics is transferred to the two point statistics when we measure the clustering of voids.

%However, these are constructed upon tetrahedra of galaxies including information on higher order statistics. While it is clear that tetrahedra are related to the 4-point statistics, they are also constrained by the 1-, 2-, and 3-point statistics, as they also depend on the mean number density of objects, the clustering of those objects, and the underlying cosmic web structure.

%The three-point function represents a measure of gravitationally induced non-Gaussianities, which characterises the morphology of the cosmic web  \citep[][]{1994ApJ...425..392F}. Moreover, the Delaunay condition selecting only empty circumspheres among those constrained by all tetrahedra, is intricately related to the cosmic web structure, and the empty spaces left after the formation of clusters and filaments. 

%In summary, the information from higher order statistics is transferred to the two point statistics when we measure the clustering of the circumsphere centres constrained by tetrahedra of galaxies. Conversely, the void 2-point correlation function from voids includes information not present in the corresponding one from galaxies.

The techniques developed in this work have directly been  used to measure the BAO from voids in observations, as presented in a companion paper (Kitaura et al.).
This work demonstrates that the clustering analysis of galaxy redshift surveys can be straightforwardly enriched by including voids.

\section*{Acknowledgments}
The authors thank {H{\'e}ctor Gil-Mar{\'i}n, Mariana Vargas-Magana, Gongbo Zhao, Juan E.~Betancort-Rijo, Gustavo Yepes, Anatoly Klypin, and Francisco Prada}  for useful discussions. YL, CZ, and CT acknowledge support by Tsinghua University with a 985 grant.
We also thank the access to computing facilities at Barcelona (MareNostrum), at LRZ (Supermuc), at AIP (erebos),  at CCIN2P3 (Quentin Le Boulc'h), and at Tsinghua University.
Funding for SDSS-III has been provided by the Alfred P. Sloan Foundation, the Participating Institutions, the National Science Foundation, and the U.S. Department of Energy Office of Science. The SDSS-III web site is http://www.sdss3.org/.

SDSS-III is managed by the Astrophysical Research Consortium for the Participating Institutions of the SDSS-III Collaboration including the University of Arizona, the Brazilian Participation Group, Brookhaven National Laboratory, Carnegie Mellon University, University of Florida, the French Participation Group, the German Participation Group, Harvard University, the Instituto de Astrofisica de Canarias, the Michigan State/Notre Dame/JINA Participation Group, Johns Hopkins University, Lawrence Berkeley National Laboratory, Max Planck Institute for Astrophysics, Max Planck Institute for Extraterrestrial Physics, New Mexico State University, New York University, Ohio State University, Pennsylvania State University, University of Portsmouth, Princeton University, the Spanish Participation Group, University of Tokyo, University of Utah, Vanderbilt University, University of Virginia, University of Washington, and Yale University.

%%%%%%%%%%%%%%%%%%%%%%%%%%%%%%%%%%%%%%%%%%%%%%%%%%



\bibliographystyle{mnras}
\bibliography{Liang_etal} % if your bibtex file is called example.bib

\begin{thebibliography}{}
\makeatletter
\relax
\def\mn@urlcharsother{\let\do\@makeother \do\$\do\&\do\#\do\^\do\_\do\%\do\~}
\def\mn@doi{\begingroup\mn@urlcharsother \@ifnextchar [ {\mn@doi@}
  {\mn@doi@[]}}
\def\mn@doi@[#1]#2{\def\@tempa{#1}\ifx\@tempa\@empty \href
  {http://dx.doi.org/#2} {doi:#2}\else \href {http://dx.doi.org/#2} {#1}\fi
  \endgroup}
\def\mn@eprint#1#2{\mn@eprint@#1:#2::\@nil}
\def\mn@eprint@arXiv#1{\href {http://arxiv.org/abs/#1} {{\tt arXiv:#1}}}
\def\mn@eprint@dblp#1{\href {http://dblp.uni-trier.de/rec/bibtex/#1.xml}
  {dblp:#1}}
\def\mn@eprint@#1:#2:#3:#4\@nil{\def\@tempa {#1}\def\@tempb {#2}\def\@tempc
  {#3}\ifx \@tempc \@empty \let \@tempc \@tempb \let \@tempb \@tempa \fi \ifx
  \@tempb \@empty \def\@tempb {arXiv}\fi \@ifundefined
  {mn@eprint@\@tempb}{\@tempb:\@tempc}{\expandafter \expandafter \csname
  mn@eprint@\@tempb\endcsname \expandafter{\@tempc}}}

\bibitem[\protect\citeauthoryear{Alam et~al.}{Alam et~al.}{2015}]{Alam:2015mbd}
Alam S.,  et~al., 2015, \mn@doi [Astrophys. J. Suppl.]
  {10.1088/0067-0049/219/1/12}, 219, 12

\bibitem[\protect\citeauthoryear{{Anderson et al.}}{{Anderson et
  al.}}{2014}]{AAB14}
{Anderson et al.} L.,  2014, \mn@doi [\mnras] {10.1093/mnras/stu523}, \href
  {http://adsabs.harvard.edu/abs/2014MNRAS.441...24A} {441, 24}

\bibitem[\protect\citeauthoryear{{Angulo}, {White}, {Springel}  \&
  {Henriques}}{{Angulo} et~al.}{2014}]{2014MNRAS.442.2131A}
{Angulo} R.~E.,  {White} S.~D.~M.,  {Springel} V.,   {Henriques} B.,  2014,
  \mn@doi [\mnras] {10.1093/mnras/stu905}, \href
  {http://adsabs.harvard.edu/abs/2014MNRAS.442.2131A} {442, 2131}

\bibitem[\protect\citeauthoryear{{Aubourg et al.}}{{Aubourg et
  al.}}{2014}]{ABB14}
{Aubourg et al.} {\'E}.,  2014, preprint, \href
  {http://adsabs.harvard.edu/abs/2014arXiv1411.1074A} {} (\mn@eprint {arXiv}
  {1411.1074})

\bibitem[\protect\citeauthoryear{{Benitez et al.}}{{Benitez et
  al.}}{2014}]{jpas2014}
{Benitez et al.} N.,  2014, preprint, \href
  {http://adsabs.harvard.edu/abs/2014arXiv1403.5237B} {} (\mn@eprint {arXiv}
  {1403.5237})

\bibitem[\protect\citeauthoryear{{Betancort-Rijo}}{{Betancort-Rijo}}{1990}]{B90}
{Betancort-Rijo} J.,  1990, \mnras, \href
  {http://adsabs.harvard.edu/abs/1990MNRAS.246..608B} {246, 608}

\bibitem[\protect\citeauthoryear{{Betancort-Rijo} \&
  {L{\'o}pez-Corredoira}}{{Betancort-Rijo} \&
  {L{\'o}pez-Corredoira}}{2002}]{BL02}
{Betancort-Rijo} J.,  {L{\'o}pez-Corredoira} M.,  2002, \mn@doi [\apj]
  {10.1086/338328}, \href {http://adsabs.harvard.edu/abs/2002ApJ...566..623B}
  {566, 623}

\bibitem[\protect\citeauthoryear{{Betancort-Rijo}, {Patiri}, {Prada}  \&
  {Romano}}{{Betancort-Rijo} et~al.}{2009}]{BPP09}
{Betancort-Rijo} J.,  {Patiri} S.~G.,  {Prada} F.,   {Romano} A.~E.,  2009,
  \mn@doi [\mnras] {10.1111/j.1365-2966.2009.15567.x}, \href
  {http://adsabs.harvard.edu/abs/2009MNRAS.400.1835B} {400, 1835}

\bibitem[\protect\citeauthoryear{{Beutler et al.}}{{Beutler et
  al.}}{2011}]{BBC11}
{Beutler et al.} F.,  2011, \mn@doi [\mnras]
  {10.1111/j.1365-2966.2011.19250.x}, \href
  {http://adsabs.harvard.edu/abs/2011MNRAS.416.3017B} {416, 3017}

\bibitem[\protect\citeauthoryear{{Beygu}, {Kreckel}, {van der Hulst},
  {Peletier}, {Jarrett}, {van de Weygaert}, {van Gorkom}  \&
  {Arag{\'o}n-Calvo}}{{Beygu} et~al.}{2015}]{BKH15}
{Beygu} B.,  {Kreckel} K.,  {van der Hulst} J.~M.,  {Peletier} R.,  {Jarrett}
  T.,  {van de Weygaert} R.,  {van Gorkom} J.~H.,   {Arag{\'o}n-Calvo} M.,
  2015, preprint, \href {http://adsabs.harvard.edu/abs/2015arXiv150102577B} {}
  (\mn@eprint {arXiv} {1501.02577})

\bibitem[\protect\citeauthoryear{{Blake et al.}}{{Blake et al.}}{2011}]{BKB11}
{Blake et al.} C.,  2011, \mn@doi [\mnras] {10.1111/j.1365-2966.2011.19592.x},
  \href {http://adsabs.harvard.edu/abs/2011MNRAS.418.1707B} {418, 1707}

\bibitem[\protect\citeauthoryear{Bolton et~al.}{Bolton
  et~al.}{2012}]{Bolton:2012hz}
Bolton A.~S.,  et~al., 2012, \mn@doi [Astron. J.]
  {10.1088/0004-6256/144/5/144}, 144, 144

\bibitem[\protect\citeauthoryear{{Bos}, {van de Weygaert}, {Dolag}  \&
  {Pettorino}}{{Bos} et~al.}{2012}]{B12}
{Bos} E.~G.~P.,  {van de Weygaert} R.,  {Dolag} K.,   {Pettorino} V.,  2012,
  \mn@doi [\mnras] {10.1111/j.1365-2966.2012.21478.x}, \href
  {http://adsabs.harvard.edu/abs/2012MNRAS.426..440B} {426, 440}

\bibitem[\protect\citeauthoryear{{Busca et al.}}{{Busca et al.}}{2013}]{BDR13}
{Busca et al.} N.~G.,  2013, \mn@doi [\aap] {10.1051/0004-6361/201220724},
  \href {http://adsabs.harvard.edu/abs/2013A%26A...552A..96B} {552, A96}

\bibitem[\protect\citeauthoryear{{Cai}, {Li}, {Cole}, {Frenk}  \&
  {Neyrinck}}{{Cai} et~al.}{2014}]{CLC14}
{Cai} Y.-C.,  {Li} B.,  {Cole} S.,  {Frenk} C.~S.,   {Neyrinck} M.,  2014,
  \mn@doi [\mnras] {10.1093/mnras/stu154}, \href
  {http://adsabs.harvard.edu/abs/2014MNRAS.439.2978C} {439, 2978}

\bibitem[\protect\citeauthoryear{{Cautun}, {van de Weygaert}  \&
  {Jones}}{{Cautun} et~al.}{2013}]{2013MNRAS.429.1286C}
{Cautun} M.,  {van de Weygaert} R.,   {Jones} B.~J.~T.,  2013, \mn@doi [\mnras]
  {10.1093/mnras/sts416}, \href
  {http://adsabs.harvard.edu/abs/2013MNRAS.429.1286C} {429, 1286}

\bibitem[\protect\citeauthoryear{{Chuang}, {Kitaura}, {Prada}, {Zhao}  \&
  {Yepes}}{{Chuang} et~al.}{2015}]{2015MNRAS.446.2621C}
{Chuang} C.-H.,  {Kitaura} F.-S.,  {Prada} F.,  {Zhao} C.,   {Yepes} G.,  2015,
  \mn@doi [\mnras] {10.1093/mnras/stu2301}, \href
  {http://adsabs.harvard.edu/abs/2015MNRAS.446.2621C} {446, 2621}

\bibitem[\protect\citeauthoryear{{Clampitt}, {Cai}  \& {Li}}{{Clampitt}
  et~al.}{2013}]{CC13}
{Clampitt} J.,  {Cai} Y.-C.,   {Li} B.,  2013, \mn@doi [\mnras]
  {10.1093/mnras/stt219}, \href
  {http://adsabs.harvard.edu/abs/2013MNRAS.431..749C} {431, 749}

\bibitem[\protect\citeauthoryear{{Clampitt}, {Jain}  \&
  {S{\'a}nchez}}{{Clampitt} et~al.}{2015}]{CJS15}
{Clampitt} J.,  {Jain} B.,   {S{\'a}nchez} C.,  2015, preprint, \href
  {http://adsabs.harvard.edu/abs/2015arXiv150708031C} {} (\mn@eprint {arXiv}
  {1507.08031})

\bibitem[\protect\citeauthoryear{{Colberg}, {Sheth}, {Diaferio}, {Gao}  \&
  {Yoshida}}{{Colberg} et~al.}{2005}]{2005MNRAS.360..216C}
{Colberg} J.~M.,  {Sheth} R.~K.,  {Diaferio} A.,  {Gao} L.,   {Yoshida} N.,
  2005, \mn@doi [\mnras] {10.1111/j.1365-2966.2005.09064.x}, \href
  {http://adsabs.harvard.edu/abs/2005MNRAS.360..216C} {360, 216}

\bibitem[\protect\citeauthoryear{{Cole et al.}}{{Cole et al.}}{2005}]{CPP05}
{Cole et al.} S.,  2005, \mn@doi [\mnras] {10.1111/j.1365-2966.2005.09318.x},
  \href {http://adsabs.harvard.edu/abs/2005MNRAS.362..505C} {362, 505}

\bibitem[\protect\citeauthoryear{{Conroy et al.}}{{Conroy et
  al.}}{2005}]{CCW05}
{Conroy et al.} C.,  2005, \mn@doi [\apj] {10.1086/497682}, \href
  {http://adsabs.harvard.edu/abs/2005ApJ...635..990C} {635, 990}

\bibitem[\protect\citeauthoryear{{Croton et al.}}{{Croton et
  al.}}{2004}]{CCG04}
{Croton et al.} D.~J.,  2004, \mn@doi [\mnras]
  {10.1111/j.1365-2966.2004.07968.x}, \href
  {http://adsabs.harvard.edu/abs/2004MNRAS.352..828C} {352, 828}

\bibitem[\protect\citeauthoryear{Dawson et~al.}{Dawson
  et~al.}{2013}]{Dawson:2012va}
Dawson K.~S.,  et~al., 2013, \mn@doi [Astron. J.] {10.1088/0004-6256/145/1/10},
  145, 10

\bibitem[\protect\citeauthoryear{{Delubac et al.}}{{Delubac et
  al.}}{2015}]{DBB15}
{Delubac et al.} T.,  2015, \mn@doi [\aap] {10.1051/0004-6361/201423969}, \href
  {http://adsabs.harvard.edu/abs/2015A%26A...574A..59D} {574, A59}

\bibitem[\protect\citeauthoryear{{Drinkwater et al.}}{{Drinkwater et
  al.}}{2010}]{wigglez2010}
{Drinkwater et al.} M.~J.,  2010, \mn@doi [\mnras]
  {10.1111/j.1365-2966.2009.15754.x}, \href
  {http://adsabs.harvard.edu/abs/2010MNRAS.401.1429D} {401, 1429}

\bibitem[\protect\citeauthoryear{{Einasto}, {Einasto}, {Gramann}  \&
  {Saar}}{{Einasto} et~al.}{1991}]{EEG91}
{Einasto} J.,  {Einasto} M.,  {Gramann} M.,   {Saar} E.,  1991, \mnras, \href
  {http://adsabs.harvard.edu/abs/1991MNRAS.248..593E} {248, 593}

\bibitem[\protect\citeauthoryear{{Eisenstein} et~al.,}{{Eisenstein}
  et~al.}{2005a}]{2005ApJ...633..560E}
{Eisenstein} D.~J.,  et~al., 2005a, \mn@doi [\apj] {10.1086/466512}, \href
  {http://adsabs.harvard.edu/abs/2005ApJ...633..560E} {633, 560}

\bibitem[\protect\citeauthoryear{{Eisenstein et al.}}{{Eisenstein et
  al.}}{2005b}]{EZH05}
{Eisenstein et al.} D.~J.,  2005b, \mn@doi [\apj] {10.1086/466512}, \href
  {http://adsabs.harvard.edu/abs/2005ApJ...633..560E} {633, 560}

\bibitem[\protect\citeauthoryear{Eisenstein et~al.}{Eisenstein
  et~al.}{2011}]{Eisenstein:2011sa}
Eisenstein D.~J.,  et~al., 2011, \mn@doi [Astron. J.]
  {10.1088/0004-6256/142/3/72}, 142, 72

\bibitem[\protect\citeauthoryear{{El-Ad} \& {Piran}}{{El-Ad} \&
  {Piran}}{1997}]{EP97}
{El-Ad} H.,  {Piran} T.,  1997, \apj, \href
  {http://adsabs.harvard.edu/abs/1997ApJ...491..421E} {491, 421}

\bibitem[\protect\citeauthoryear{{Forero-Romero}, {Hoffman}, {Gottl{\"o}ber},
  {Klypin}  \& {Yepes}}{{Forero-Romero} et~al.}{2009}]{2009MNRAS.396.1815F}
{Forero-Romero} J.~E.,  {Hoffman} Y.,  {Gottl{\"o}ber} S.,  {Klypin} A.,
  {Yepes} G.,  2009, \mn@doi [\mnras] {10.1111/j.1365-2966.2009.14885.x}, \href
  {http://adsabs.harvard.edu/abs/2009MNRAS.396.1815F} {396, 1815}

\bibitem[\protect\citeauthoryear{{Frieman} \& {Dark Energy Survey
  Collaboration}}{{Frieman} \& {Dark Energy Survey
  Collaboration}}{2013}]{des2013}
{Frieman} J.,  {Dark Energy Survey Collaboration} 2013, in American
  Astronomical Society Meeting Abstracts. p. 335.01

\bibitem[\protect\citeauthoryear{Fukugita, Ichikawa, Gunn, Doi, Shimasaku  \&
  Schneider}{Fukugita et~al.}{1996}]{Fukugita:1996qt}
Fukugita M.,  Ichikawa T.,  Gunn J.~E.,  Doi M.,  Shimasaku K.,   Schneider
  D.~P.,  1996, \mn@doi [Astron. J.] {10.1086/117915}, 111, 1748

\bibitem[\protect\citeauthoryear{{Granett}, {Neyrinck}  \& {Szapudi}}{{Granett}
  et~al.}{2008}]{GNS08}
{Granett} B.~R.,  {Neyrinck} M.~C.,   {Szapudi} I.,  2008, \mn@doi [\apjl]
  {10.1086/591670}, \href {http://adsabs.harvard.edu/abs/2008ApJ...683L..99G}
  {683, L99}

\bibitem[\protect\citeauthoryear{Gunn et~al.}{Gunn et~al.}{1998}]{Gunn:1998vh}
Gunn J.~E.,  et~al., 1998, \mn@doi [Astron. J.] {10.1086/300645}, 116, 3040

\bibitem[\protect\citeauthoryear{Gunn et~al.}{Gunn et~al.}{2006}]{Gunn:2006tw}
Gunn J.~E.,  et~al., 2006, \mn@doi [Astron. J.] {10.1086/500975}, 131, 2332

\bibitem[\protect\citeauthoryear{{Hahn}, {Porciani}, {Carollo}  \&
  {Dekel}}{{Hahn} et~al.}{2007}]{2007MNRAS.375..489H}
{Hahn} O.,  {Porciani} C.,  {Carollo} C.~M.,   {Dekel} A.,  2007, \mn@doi
  [\mnras] {10.1111/j.1365-2966.2006.11318.x}, \href
  {http://adsabs.harvard.edu/abs/2007MNRAS.375..489H} {375, 489}

\bibitem[\protect\citeauthoryear{Hamaus, Wandelt, Sutter, Lavaux  \&
  Warren}{Hamaus et~al.}{2014}]{Hamaus:2013qja}
Hamaus N.,  Wandelt B.~D.,  Sutter P.~M.,  Lavaux G.,   Warren M.~S.,  2014,
  \mn@doi [Phys. Rev. Lett.] {10.1103/PhysRevLett.112.041304}, 112, 041304

\bibitem[\protect\citeauthoryear{{Hinshaw et al.}}{{Hinshaw et
  al.}}{2013}]{WMAP913}
{Hinshaw et al.} G.,  2013, \mn@doi [\apjs] {10.1088/0067-0049/208/2/19}, \href
  {http://adsabs.harvard.edu/abs/2013ApJS..208...19H} {208, 19}

\bibitem[\protect\citeauthoryear{{Hoffman}, {Metuki}, {Yepes}, {Gottl{\"o}ber},
  {Forero-Romero}, {Libeskind}  \& {Knebe}}{{Hoffman}
  et~al.}{2012}]{2012MNRAS.425.2049H}
{Hoffman} Y.,  {Metuki} O.,  {Yepes} G.,  {Gottl{\"o}ber} S.,  {Forero-Romero}
  J.~E.,  {Libeskind} N.~I.,   {Knebe} A.,  2012, \mn@doi [\mnras]
  {10.1111/j.1365-2966.2012.21553.x}, \href
  {http://adsabs.harvard.edu/abs/2012MNRAS.425.2049H} {425, 2049}

\bibitem[\protect\citeauthoryear{{Hotchkiss}, {Nadathur}, {Gottl{\"o}ber},
  {Iliev}, {Knebe}, {Watson}  \& {Yepes}}{{Hotchkiss} et~al.}{2015}]{HNG15}
{Hotchkiss} S.,  {Nadathur} S.,  {Gottl{\"o}ber} S.,  {Iliev} I.~T.,  {Knebe}
  A.,  {Watson} W.~A.,   {Yepes} G.,  2015, \mn@doi [\mnras]
  {10.1093/mnras/stu2072}, \href
  {http://adsabs.harvard.edu/abs/2015MNRAS.446.1321H} {446, 1321}

\bibitem[\protect\citeauthoryear{{Hoyle} \& {Vogeley}}{{Hoyle} \&
  {Vogeley}}{2004}]{HV04}
{Hoyle} F.,  {Vogeley} M.~S.,  2004, \mn@doi [\apj] {10.1086/386279}, \href
  {http://adsabs.harvard.edu/abs/2004ApJ...607..751H} {607, 751}

\bibitem[\protect\citeauthoryear{{Ili{\'c}}, {Langer}  \& {Douspis}}{{Ili{\'c}}
  et~al.}{2013}]{ILD13}
{Ili{\'c}} S.,  {Langer} M.,   {Douspis} M.,  2013, \mn@doi [\aap]
  {10.1051/0004-6361/201321150}, \href
  {http://adsabs.harvard.edu/abs/2013A%26A...556A..51I} {556, A51}

\bibitem[\protect\citeauthoryear{{Jasche}, {Kitaura}, {Li}  \&
  {En{\ss}lin}}{{Jasche} et~al.}{2010}]{JKL10}
{Jasche} J.,  {Kitaura} F.~S.,  {Li} C.,   {En{\ss}lin} T.~A.,  2010, \mn@doi
  [\mnras] {10.1111/j.1365-2966.2010.17313.x}, \href
  {http://adsabs.harvard.edu/abs/2010MNRAS.tmp.1638J} {pp 1638--+}

\bibitem[\protect\citeauthoryear{{Kirshner}, {Oemler}, {Schechter}  \&
  {Shectman}}{{Kirshner} et~al.}{1981}]{KOS81}
{Kirshner} R.~P.,  {Oemler} Jr. A.,  {Schechter} P.~L.,   {Shectman} S.~A.,
  1981, \mn@doi [\apjl] {10.1086/183623}, \href
  {http://adsabs.harvard.edu/abs/1981ApJ...248L..57K} {248, L57}

\bibitem[\protect\citeauthoryear{{Kitaura}, {Jasche}, {Li}, {En{\ss}lin},
  {Metcalf}, {Wandelt}, {Lemson}  \& {White}}{{Kitaura} et~al.}{2009}]{K09}
{Kitaura} F.~S.,  {Jasche} J.,  {Li} C.,  {En{\ss}lin} T.~A.,  {Metcalf} R.~B.,
   {Wandelt} B.~D.,  {Lemson} G.,   {White} S.~D.~M.,  2009, \mn@doi [\mnras]
  {10.1111/j.1365-2966.2009.15470.x}, \href
  {http://adsabs.harvard.edu/abs/2009MNRAS.400..183K} {400, 183}

\bibitem[\protect\citeauthoryear{{Kitaura}, {Yepes}  \& {Prada}}{{Kitaura}
  et~al.}{2014}]{2014MNRAS.439L..21K}
{Kitaura} F.-S.,  {Yepes} G.,   {Prada} F.,  2014, \mn@doi [\mnras]
  {10.1093/mnrasl/slt172}, \href
  {http://adsabs.harvard.edu/abs/2014MNRAS.439L..21K} {439, L21}

\bibitem[\protect\citeauthoryear{{Kitaura} et~al.,}{{Kitaura}
  et~al.}{2015a}]{2015arXiv150906400K}
{Kitaura} F.-S.,  et~al., 2015a, preprint, \href
  {http://adsabs.harvard.edu/abs/2015arXiv150906400K} {} (\mn@eprint {arXiv}
  {1509.06400})

\bibitem[\protect\citeauthoryear{{Kitaura}, {Gil-Mar{\'{\i}}n}, {Sc{\'o}ccola},
  {Chuang}, {M{\"u}ller}, {Yepes}  \& {Prada}}{{Kitaura}
  et~al.}{2015b}]{2015MNRAS.450.1836K}
{Kitaura} F.-S.,  {Gil-Mar{\'{\i}}n} H.,  {Sc{\'o}ccola} C.~G.,  {Chuang}
  C.-H.,  {M{\"u}ller} V.,  {Yepes} G.,   {Prada} F.,  2015b, \mn@doi [\mnras]
  {10.1093/mnras/stv645}, \href
  {http://adsabs.harvard.edu/abs/2015MNRAS.450.1836K} {450, 1836}

\bibitem[\protect\citeauthoryear{{Klypin} \& {Rhee}}{{Klypin} \&
  {Rhee}}{1994}]{1994ApJ...428..399K}
{Klypin} A.,  {Rhee} G.,  1994, \mn@doi [\apj] {10.1086/174252}, \href
  {http://adsabs.harvard.edu/abs/1994ApJ...428..399K} {428, 399}

\bibitem[\protect\citeauthoryear{{Klypin}, {Yepes}, {Gottlober}, {Prada}  \&
  {Hess}}{{Klypin} et~al.}{2014}]{Klypin2014}
{Klypin} A.,  {Yepes} G.,  {Gottlober} S.,  {Prada} F.,   {Hess} S.,  2014,
  preprint, \href {http://adsabs.harvard.edu/abs/2014arXiv1411.4001K} {}
  (\mn@eprint {arXiv} {1411.4001})

\bibitem[\protect\citeauthoryear{{Komatsu et al.}}{{Komatsu et
  al.}}{2011}]{WMAP711}
{Komatsu et al.} E.,  2011, \mn@doi [\apjs] {10.1088/0067-0049/192/2/18}, \href
  {http://adsabs.harvard.edu/abs/2011ApJS..192...18K} {192, 18}

\bibitem[\protect\citeauthoryear{{LSST Dark Energy Science
  Collaboration}}{{LSST Dark Energy Science Collaboration}}{2012}]{lsst2012}
{LSST Dark Energy Science Collaboration} 2012, preprint, \href
  {http://adsabs.harvard.edu/abs/2012arXiv1211.0310L} {} (\mn@eprint {arXiv}
  {1211.0310})

\bibitem[\protect\citeauthoryear{{Lam}, {Clampitt}, {Cai}  \& {Li}}{{Lam}
  et~al.}{2015}]{LCC15}
{Lam} T.~Y.,  {Clampitt} J.,  {Cai} Y.-C.,   {Li} B.,  2015, \mn@doi [\mnras]
  {10.1093/mnras/stv797}, \href
  {http://adsabs.harvard.edu/abs/2015MNRAS.450.3319L} {450, 3319}

\bibitem[\protect\citeauthoryear{{Landy} \& {Szalay}}{{Landy} \&
  {Szalay}}{1993}]{1993ApJ...412...64L}
{Landy} S.~D.,  {Szalay} A.~S.,  1993, \mn@doi [\apj] {10.1086/172900}, \href
  {http://adsabs.harvard.edu/abs/1993ApJ...412...64L} {412, 64}

\bibitem[\protect\citeauthoryear{{Laureijs}}{{Laureijs}}{2009}]{euclid2009}
{Laureijs} R.,  2009, preprint, \href
  {http://adsabs.harvard.edu/abs/2009arXiv0912.0914L} {} (\mn@eprint {arXiv}
  {0912.0914})

\bibitem[\protect\citeauthoryear{{Lavaux} \& {Wandelt}}{{Lavaux} \&
  {Wandelt}}{2010}]{LW10}
{Lavaux} G.,  {Wandelt} B.~D.,  2010, \mn@doi [\mnras]
  {10.1111/j.1365-2966.2010.16197.x}, \href
  {http://adsabs.harvard.edu/abs/2010MNRAS.403.1392L} {403, 1392}

\bibitem[\protect\citeauthoryear{{Li}}{{Li}}{2011}]{L11}
{Li} B.,  2011, \mn@doi [\mnras] {10.1111/j.1365-2966.2010.17867.x}, \href
  {http://adsabs.harvard.edu/abs/2011MNRAS.411.2615L} {411, 2615}

\bibitem[\protect\citeauthoryear{{Li}, {Zhao}  \& {Koyama}}{{Li}
  et~al.}{2012}]{LZK12}
{Li} B.,  {Zhao} G.-B.,   {Koyama} K.,  2012, \mn@doi [\mnras]
  {10.1111/j.1365-2966.2012.20573.x}, \href
  {http://adsabs.harvard.edu/abs/2012MNRAS.421.3481L} {421, 3481}

\bibitem[\protect\citeauthoryear{{Martino} \& {Sheth}}{{Martino} \&
  {Sheth}}{2009}]{MS09}
{Martino} M.~C.,  {Sheth} R.~K.,  2009, preprint, \href
  {http://adsabs.harvard.edu/abs/2009arXiv0911.1829M} {} (\mn@eprint {arXiv}
  {0911.1829})

\bibitem[\protect\citeauthoryear{{M{\"u}ller}, {Arbabi-Bidgoli}, {Einasto}  \&
  {Tucker}}{{M{\"u}ller} et~al.}{2000}]{MAE00}
{M{\"u}ller} V.,  {Arbabi-Bidgoli} S.,  {Einasto} J.,   {Tucker} D.,  2000,
  \mn@doi [\mnras] {10.1046/j.1365-8711.2000.03775.x}, \href
  {http://adsabs.harvard.edu/abs/2000MNRAS.318..280M} {318, 280}

\bibitem[\protect\citeauthoryear{{Nadathur} \& {Hotchkiss}}{{Nadathur} \&
  {Hotchkiss}}{2014}]{NH14}
{Nadathur} S.,  {Hotchkiss} S.,  2014, \mn@doi [\mnras] {10.1093/mnras/stu349},
  \href {http://adsabs.harvard.edu/abs/2014MNRAS.440.1248N} {440, 1248}

\bibitem[\protect\citeauthoryear{{Neyrinck}}{{Neyrinck}}{2008}]{2008MNRAS.386.2101N}
{Neyrinck} M.~C.,  2008, \mn@doi [\mnras] {10.1111/j.1365-2966.2008.13180.x},
  \href {http://adsabs.harvard.edu/abs/2008MNRAS.386.2101N} {386, 2101}

\bibitem[\protect\citeauthoryear{{Nuza}, {Kitaura}, {He{\ss}}, {Libeskind}  \&
  {M{\"u}ller}}{{Nuza} et~al.}{2014}]{NKH14}
{Nuza} S.~E.,  {Kitaura} F.-S.,  {He{\ss}} S.,  {Libeskind} N.~I.,
  {M{\"u}ller} V.,  2014, \mn@doi [\mnras] {10.1093/mnras/stu1746}, \href
  {http://adsabs.harvard.edu/abs/2014MNRAS.445..988N} {445, 988}

\bibitem[\protect\citeauthoryear{{Pan}, {Vogeley}, {Hoyle}, {Choi}  \&
  {Park}}{{Pan} et~al.}{2012}]{PVH12}
{Pan} D.~C.,  {Vogeley} M.~S.,  {Hoyle} F.,  {Choi} Y.-Y.,   {Park} C.,  2012,
  \mn@doi [\mnras] {10.1111/j.1365-2966.2011.20197.x}, \href
  {http://adsabs.harvard.edu/abs/2012MNRAS.421..926P} {421, 926}

\bibitem[\protect\citeauthoryear{{Park} \& {Lee}}{{Park} \& {Lee}}{2007}]{PL07}
{Park} D.,  {Lee} J.,  2007, \mn@doi [Physical Review Letters]
  {10.1103/PhysRevLett.98.081301}, \href
  {http://adsabs.harvard.edu/abs/2007PhRvL..98h1301P} {98, 081301}

\bibitem[\protect\citeauthoryear{{Patiri}, {Betancort-Rijo}, {Prada}, {Klypin}
  \& {Gottl{\"o}ber}}{{Patiri} et~al.}{2006a}]{PBP06}
{Patiri} S.~G.,  {Betancort-Rijo} J.~E.,  {Prada} F.,  {Klypin} A.,
  {Gottl{\"o}ber} S.,  2006a, \mn@doi [\mnras]
  {10.1111/j.1365-2966.2006.10305.x}, \href
  {http://adsabs.harvard.edu/abs/2006MNRAS.369..335P} {369, 335}

\bibitem[\protect\citeauthoryear{{Patiri}, {Prada}, {Holtzman}, {Klypin}  \&
  {Betancort-Rijo}}{{Patiri} et~al.}{2006b}]{PPH06}
{Patiri} S.~G.,  {Prada} F.,  {Holtzman} J.,  {Klypin} A.,   {Betancort-Rijo}
  J.,  2006b, \mn@doi [\mnras] {10.1111/j.1365-2966.2006.10975.x}, \href
  {http://adsabs.harvard.edu/abs/2006MNRAS.372.1710P} {372, 1710}

\bibitem[\protect\citeauthoryear{{Peebles} \& {Hauser}}{{Peebles} \&
  {Hauser}}{1974}]{1974ApJS...28...19P}
{Peebles} P.~J.~E.,  {Hauser} M.~G.,  1974, \mn@doi [\apjs] {10.1086/190308},
  \href {http://adsabs.harvard.edu/abs/1974ApJS...28...19P} {28, 19}

\bibitem[\protect\citeauthoryear{{Percival et al.}}{{Percival et
  al.}}{2010}]{PRE10}
{Percival et al.} W.~J.,  2010, \mn@doi [\mnras]
  {10.1111/j.1365-2966.2009.15812.x}, \href
  {http://adsabs.harvard.edu/abs/2010MNRAS.401.2148P} {401, 2148}

\bibitem[\protect\citeauthoryear{{Perlmutter} et~al.,}{{Perlmutter}
  et~al.}{1998}]{1998Natur.391...51P}
{Perlmutter} S.,  et~al., 1998, \mn@doi [\nat] {10.1038/34124}, \href
  {http://adsabs.harvard.edu/abs/1998Natur.391...51P} {391, 51}

\bibitem[\protect\citeauthoryear{{Pisani}, {Sutter}, {Hamaus}, {Alizadeh},
  {Biswas}, {Wandelt}  \& {Hirata}}{{Pisani}
  et~al.}{2015}]{2015arXiv150307690P}
{Pisani} A.,  {Sutter} P.~M.,  {Hamaus} N.,  {Alizadeh} E.,  {Biswas} R.,
  {Wandelt} B.~D.,   {Hirata} C.~M.,  2015, preprint, \href
  {http://adsabs.harvard.edu/abs/2015arXiv150307690P} {} (\mn@eprint {arXiv}
  {1503.07690})

\bibitem[\protect\citeauthoryear{{Planck Collaboration}}{{Planck
  Collaboration}}{2014a}]{PLANCKBAO14}
{Planck Collaboration} 2014a, \mn@doi [\aap] {10.1051/0004-6361/201321529},
  \href {http://adsabs.harvard.edu/abs/2014A%26A...571A...1P} {571, A1}

\bibitem[\protect\citeauthoryear{{Planck Collaboration}}{{Planck
  Collaboration}}{2014b}]{PLANCKISW14}
{Planck Collaboration} 2014b, \mn@doi [\aap] {10.1051/0004-6361/201321526},
  \href {http://adsabs.harvard.edu/abs/2014A%26A...571A..19P} {571, A19}

\bibitem[\protect\citeauthoryear{{Platen}, {van de Weygaert}  \&
  {Jones}}{{Platen} et~al.}{2007}]{2007MNRAS.380..551P}
{Platen} E.,  {van de Weygaert} R.,   {Jones} B.~J.~T.,  2007, \mn@doi [\mnras]
  {10.1111/j.1365-2966.2007.12125.x}, \href
  {http://adsabs.harvard.edu/abs/2007MNRAS.380..551P} {380, 551}

\bibitem[\protect\citeauthoryear{{Platen}, {van de Weygaert}, {Jones}, {Vegter}
   \& {Calvo}}{{Platen} et~al.}{2011}]{PWJ11}
{Platen} E.,  {van de Weygaert} R.,  {Jones} B.~J.~T.,  {Vegter} G.,   {Calvo}
  M.~A.~A.,  2011, \mn@doi [\mnras] {10.1111/j.1365-2966.2011.18905.x}, \href
  {http://adsabs.harvard.edu/abs/2011MNRAS.416.2494P} {416, 2494}

\bibitem[\protect\citeauthoryear{{Plionis} \& {Basilakos}}{{Plionis} \&
  {Basilakos}}{2002}]{PB02}
{Plionis} M.,  {Basilakos} S.,  2002, \mn@doi [\mnras]
  {10.1046/j.1365-8711.2002.05069.x}, \href
  {http://adsabs.harvard.edu/abs/2002MNRAS.330..399P} {330, 399}

\bibitem[\protect\citeauthoryear{{Politzer} \& {Preskill}}{{Politzer} \&
  {Preskill}}{1986}]{PP86}
{Politzer} H.~D.,  {Preskill} J.~P.,  1986, \mn@doi [Physical Review Letters]
  {10.1103/PhysRevLett.56.99}, \href
  {http://adsabs.harvard.edu/abs/1986PhRvL..56...99P} {56, 99}

\bibitem[\protect\citeauthoryear{{Prada}, {Klypin}, {Yepes}, {Nuza}  \&
  {Gottloeber}}{{Prada} et~al.}{2011}]{2011arXiv1111.2889P}
{Prada} F.,  {Klypin} A.,  {Yepes} G.,  {Nuza} S.~E.,   {Gottloeber} S.,  2011,
  preprint, \href {http://adsabs.harvard.edu/abs/2011arXiv1111.2889P} {}
  (\mn@eprint {arXiv} {1111.2889})

\bibitem[\protect\citeauthoryear{{Prada}, {Scoccola}, {Chuang}, {Yepes},
  {Klypin}, {Kitaura}, {Gottlober}  \& {Zhao}}{{Prada}
  et~al.}{2014}]{2014arXiv1410.4684P}
{Prada} F.,  {Scoccola} C.~G.,  {Chuang} C.-H.,  {Yepes} G.,  {Klypin} A.~A.,
  {Kitaura} F.-S.,  {Gottlober} S.,   {Zhao} C.,  2014, preprint, \href
  {http://adsabs.harvard.edu/abs/2014arXiv1410.4684P} {} (\mn@eprint {arXiv}
  {1410.4684})

\bibitem[\protect\citeauthoryear{Reid et~al.}{Reid et~al.}{2015}]{Reid:2015gra}
Reid B.,  et~al., 2015, preprint (\mn@eprint {arXiv} {1509.06529})

\bibitem[\protect\citeauthoryear{{Riess} et~al.,}{{Riess}
  et~al.}{1998}]{1998AJ....116.1009R}
{Riess} A.~G.,  et~al., 1998, \mn@doi [\aj] {10.1086/300499}, \href
  {http://adsabs.harvard.edu/abs/1998AJ....116.1009R} {116, 1009}

\bibitem[\protect\citeauthoryear{{Rodr{\'{\i}}guez-Torres}
  et~al.,}{{Rodr{\'{\i}}guez-Torres} et~al.}{2015}]{2015arXiv150906404R}
{Rodr{\'{\i}}guez-Torres} S.~A.,  et~al., 2015, preprint, \href
  {http://adsabs.harvard.edu/abs/2015arXiv150906404R} {} (\mn@eprint {arXiv}
  {1509.06404})

\bibitem[\protect\citeauthoryear{{Schlegel et al.}}{{Schlegel et
  al.}}{2011}]{bigboss2011}
{Schlegel et al.} D.,  2011, preprint, \href
  {http://adsabs.harvard.edu/abs/2011arXiv1106.1706S} {} (\mn@eprint {arXiv}
  {1106.1706})

\bibitem[\protect\citeauthoryear{{Schmidt} et~al.,}{{Schmidt}
  et~al.}{1998}]{1998ApJ...507...46S}
{Schmidt} B.~P.,  et~al., 1998, \mn@doi [\apj] {10.1086/306308}, \href
  {http://adsabs.harvard.edu/abs/1998ApJ...507...46S} {507, 46}

\bibitem[\protect\citeauthoryear{{Shandarin}, {Feldman}, {Heitmann}  \&
  {Habib}}{{Shandarin} et~al.}{2006}]{2006MNRAS.367.1629S}
{Shandarin} S.,  {Feldman} H.~A.,  {Heitmann} K.,   {Habib} S.,  2006, \mn@doi
  [\mnras] {10.1111/j.1365-2966.2006.10062.x}, \href
  {http://adsabs.harvard.edu/abs/2006MNRAS.367.1629S} {367, 1629}

\bibitem[\protect\citeauthoryear{{Sheth} \& {van de Weygaert}}{{Sheth} \& {van
  de Weygaert}}{2004}]{SW04}
{Sheth} R.~K.,  {van de Weygaert} R.,  2004, \mn@doi [\mnras]
  {10.1111/j.1365-2966.2004.07661.x}, \href
  {http://adsabs.harvard.edu/abs/2004MNRAS.350..517S} {350, 517}

\bibitem[\protect\citeauthoryear{{Slosar et al.}}{{Slosar et
  al.}}{2013}]{SIK13}
{Slosar et al.} A.,  2013, \mn@doi [\jcap] {10.1088/1475-7516/2013/04/026},
  \href {http://adsabs.harvard.edu/abs/2013JCAP...04..026S} {4, 26}

\bibitem[\protect\citeauthoryear{Smee et~al.}{Smee et~al.}{2013}]{Smee:2012wd}
Smee S.,  et~al., 2013, \mn@doi [Astron. J.] {10.1088/0004-6256/146/2/32}, 146,
  32

\bibitem[\protect\citeauthoryear{{Spergel et al.}}{{Spergel et
  al.}}{2003}]{WMAP103}
{Spergel et al.} D.~N.,  2003, \mn@doi [\apjs] {10.1086/377226}, \href
  {http://adsabs.harvard.edu/abs/2003ApJS..148..175S} {148, 175}

\bibitem[\protect\citeauthoryear{{Sutter}, {Lavaux}, {Wandelt}  \&
  {Weinberg}}{{Sutter} et~al.}{2012}]{SLW12}
{Sutter} P.~M.,  {Lavaux} G.,  {Wandelt} B.~D.,   {Weinberg} D.~H.,  2012,
  \mn@doi [\apj] {10.1088/0004-637X/761/2/187}, \href
  {http://adsabs.harvard.edu/abs/2012ApJ...761..187S} {761, 187}

\bibitem[\protect\citeauthoryear{{Sutter}, {Lavaux}, {Wandelt}, {Weinberg},
  {Warren}  \& {Pisani}}{{Sutter} et~al.}{2014}]{SLW14}
{Sutter} P.~M.,  {Lavaux} G.,  {Wandelt} B.~D.,  {Weinberg} D.~H.,  {Warren}
  M.~S.,   {Pisani} A.,  2014, \mn@doi [\mnras] {10.1093/mnras/stu1094}, \href
  {http://adsabs.harvard.edu/abs/2014MNRAS.442.3127S} {442, 3127}

\bibitem[\protect\citeauthoryear{{Sylos Labini}, {Vasilyev}, {Baryshev}  \&
  {L{\'o}pez-Corredoira}}{{Sylos Labini} et~al.}{2009}]{2009A&A...505..981S}
{Sylos Labini} F.,  {Vasilyev} N.~L.,  {Baryshev} Y.~V.,
  {L{\'o}pez-Corredoira} M.,  2009, \mn@doi [\aap]
  {10.1051/0004-6361/200911987}, \href
  {http://adsabs.harvard.edu/abs/2009A%26A...505..981S} {505, 981}

\bibitem[\protect\citeauthoryear{{The CGAL Project}}{{The CGAL
  Project}}{2015}]{cgal:eb-15b}
{The CGAL Project} 2015, {CGAL} User and Reference Manual, {4.6.3} edn.
{CGAL Editorial Board}, \url {http://doc.cgal.org/4.7/Manual/packages.html}

\bibitem[\protect\citeauthoryear{{Varela}, {Betancort-Rijo}, {Trujillo}  \&
  {Ricciardelli}}{{Varela} et~al.}{2012}]{VBT12}
{Varela} J.,  {Betancort-Rijo} J.,  {Trujillo} I.,   {Ricciardelli} E.,  2012,
  \mn@doi [\apj] {10.1088/0004-637X/744/2/82}, \href
  {http://adsabs.harvard.edu/abs/2012ApJ...744...82V} {744, 82}

\bibitem[\protect\citeauthoryear{{Vogeley}, {Geller}, {Park}  \&
  {Huchra}}{{Vogeley} et~al.}{1994}]{VGP94}
{Vogeley} M.~S.,  {Geller} M.~J.,  {Park} C.,   {Huchra} J.~P.,  1994, \mn@doi
  [\aj] {10.1086/117110}, \href
  {http://adsabs.harvard.edu/abs/1994AJ....108..745V} {108, 745}

\bibitem[\protect\citeauthoryear{{White}}{{White}}{1979}]{W79}
{White} S.~D.~M.,  1979, \mnras, \href
  {http://adsabs.harvard.edu/abs/1979MNRAS.186..145W} {186, 145}

\bibitem[\protect\citeauthoryear{{White et al.}}{{White et
  al.}}{2011}]{boss2011}
{White et al.} M.,  2011, \mn@doi [\apj] {10.1088/0004-637X/728/2/126}, \href
  {http://adsabs.harvard.edu/abs/2011ApJ...728..126W} {728, 126}

\bibitem[\protect\citeauthoryear{{White}, {Tinker}  \& {McBride}}{{White}
  et~al.}{2014}]{2014MNRAS.437.2594W}
{White} M.,  {Tinker} J.~L.,   {McBride} C.~K.,  2014, \mn@doi [\mnras]
  {10.1093/mnras/stt2071}, \href
  {http://adsabs.harvard.edu/abs/2014MNRAS.437.2594W} {437, 2594}

\bibitem[\protect\citeauthoryear{York et~al.}{York et~al.}{2000}]{York:2000gk}
York D.~G.,  et~al., 2000, \mn@doi [Astron. J.] {10.1086/301513}, 120, 1579

\bibitem[\protect\citeauthoryear{{de Jong et al.}}{{de Jong et
  al.}}{2012}]{4most}
{de Jong et al.} R.~S.,  2012, in Society of Photo-Optical Instrumentation
  Engineers (SPIE) Conference Series.  (\mn@eprint {arXiv} {1206.6885}),
  \mn@doi{10.1117/12.926239}

\bibitem[\protect\citeauthoryear{{de Lapparent}, {Geller}  \& {Huchra}}{{de
  Lapparent} et~al.}{1986}]{LGH86}
{de Lapparent} V.,  {Geller} M.~J.,   {Huchra} J.~P.,  1986, \mn@doi [\apjl]
  {10.1086/184625}, \href {http://adsabs.harvard.edu/abs/1986ApJ...302L...1D}
  {302, L1}

\makeatother
\end{thebibliography}

%\begin{thebibliography}{99}
%\end{thebibliography}

%%%%%%%%%%%%%%%%%%%%%%%%%%%%%%%%%%%%%%%%%%%%%%%%%%

%%%%%%%%%%%%%%%%% APPENDICES %%%%%%%%%%%%%%%%%%%%%
%\appendix

%%%%%%%%%%%%%%%%%%%%%%%%%%%%%%%%%%%%%%%%%%%%%%%%%%

% Don't change these lines
\bsp	% typesetting comment
\label{lastpage}
\end{document}